\documentclass[aps,prx,twocolumn,floatfix]{revtex4}
\usepackage{latexsym}

\usepackage{subfigure}    
\usepackage{epsfig}

\usepackage{graphicx}
\usepackage{dcolumn}
\usepackage{bm}
\usepackage{epstopdf}
\usepackage{natbib}
\usepackage{bm}
\usepackage{color}

\newcommand{\figwidth}{2.0 in}

\newcommand{\figwidthc}{2.8 in}
\newcommand{\figwc}{2.6 in}
\newcommand{\figwidthsmall}{1.7 in}

\usepackage{subfigure}
\usepackage{hyperref}

\begin{document}


\title{Converting topological insulators into topological metals within the tetradymite family}

\author{K. -W. Chen$^{1,2}$, N. Aryal$^{1,2}$, J. Dai$^{3}$, D. Graf$^{1}$, S. Zhang$^{1,2}$, S. Das$^{1,2}$, P.~Le~F\`evre$^{5}$, F.~Bertran$^{5}$, R.~Yukawa$^{4}$, K.~Horiba$^{4}$, H.~Kumigashira$^{4}$, E. Frantzeskakis$^{3}$,  F. Fortuna$^{3}$, L. Balicas$^{1,2}$, A. F. Santander-Syro$^{3}$, E. Manousakis$^{1,2}$, and R. E. Baumbach$^{1,2}$}
\affiliation{%
 $^1$National High Magnetic Field Laboratory, Florida State University\\
 $^2$Department of Physics, Florida State University\\
 $^3$CSNSM, Univ. Paris-Sud, CNRS/IN2P3, Universit\'e Paris-Saclay, 91405 Orsay Cedex, France\\
 $^4$Photon Factory, Institute of Materials Structure Science, High Energy Accelerator Research Organization (KEK), 1-1 Oho, Tsukuba 305-0801, Japan\\
 $^5$Synchrotron SOLEIL, L'Orme des Merisiers, Saint-Aubin-BP48, 91192 Gif-sur-Yvette, France
}%








\begin{abstract}
We report the electronic band structures and concomitant Fermi surfaces for a family of exfoliable tetradymite compounds with the formula $T_2$$Ch_2$$Pn$, obtained as a modification to the well-known topological insulator binaries Bi$_2$(Se,Te)$_3$ by replacing one chalcogen ($Ch$) with a pnictogen ($Pn$) and Bi with the tetravalent transition metals $T$ $=$ Ti, Zr, or Hf. This imbalances the electron count and results in layered metals characterized by relatively high carrier mobilities and bulk two-dimensional Fermi surfaces whose topography is well-described by first principles calculations. Intriguingly, slab electronic structure calculations predict Dirac-like surface states. In contrast to Bi$_2$Se$_3$, where the surface Dirac bands are at the $\Gamma-$point, for (Zr,Hf)$_2$Te$_2$(P,As) there are Dirac cones of strong topological character around both the $\bar {\Gamma}$- and $\bar {M}$-points which are above and below the Fermi energy, respectively. For Ti$_2$Te$_2$P the surface state is predicted to exist only around the $\bar {M}$-point. In agreement with these predictions, the surface states that are located below the Fermi energy are observed by angle resolved photoemission spectroscopy measurements, revealing that they coexist with the bulk metallic state. Thus, this family of materials provides a foundation upon which to develop novel phenomena that exploit both the bulk and surface states (e.g., topological superconductivity).
\end{abstract}

\maketitle


\section{Introduction}
The compounds Bi$_2$Se$_3$ and Bi$_2$Te$_3$, which crystallize in the same structure as the naturally occurring mineral ``tetradymite" (Bi$_2$Te$_2$S~\cite{Wang2011,harker,bos}),
have attracted attention as the first examples of three dimensional topological insulators~\cite{Zhang2009, Chen178,Hsieh2009,Xia2009}. These and related materials are of interest
because their electronic state is gapped in the bulk as in band insulators, but has Dirac-like surface state dispersions which are protected
by particle number conservation and time reversal symmetry~\cite{kane, wen,pollmann,wen2}. Hence, their electrical transport properties are dictated by the topology of the
electronic bands of their surface state instead of a broken symmetry. This leads to protection against a variety of common scattering mechanisms and electrical properties which are distinct from those of typical metals or insulators ~\cite{Geim2007,Geim2009,Neto2009,Ando2013,Hasan2010}. Besides topological insulators, a variety of other semimetallic systems exhibit novel behavior resulting from the interplay between symmetry and topology. Well known examples include the Dirac and Weyl semimetals \cite{liu2014,neupane2014,xu2015,soluyanov2015}, although the broader family of topological materials continues to expand~\cite{xu2015a,bian2016,huang2016}.

An important feature in such materials is whether or not there are tunable parameters to optimize properties: e.g., by varying i) the nearness of the Fermi level to a Dirac point or ii) the strength of the spin orbit coupling to open or close a gap at a Dirac point. In this context, the tetradymite family is a useful template because it can be intercalated with charge 
dopants~\cite{hor2010,asaba2017} and also hosts pronounced chemical/structural flexibility. For instance, the minerals ``pilsenite'' (Bi$_4$Te$_3$~\cite{jeffries}) and ``tsumoite'' (BiTe~\cite{yamana}) are stacking variants within the generalized formula (Bi$_2$)$_m$(Bi$_2$Te$_3$)$_n$~\cite{bos} where Bi$_2$Te$_2$Se, Bi$_2$Te$_2$S, Bi$_2$Se$_2$S, Sb$_2$Te$_2$Se and Sb$_2$Te$_2$S are ternary charge balanced variants~\cite{Lin2011,Wang2011,ren10,wang_14,reimann_14}. Even elemental bismuth, which exhibits anomalous semimetallic behavior \cite{behnia_Bi}, is a member of this family. More recently, Zr$_2$Te$_2$P was introduced as a strong topological metal with multiple Dirac cones ~\cite{Ji_16,chen_16_2}.
Zr$_2$Te$_2$P is electronically distinct from the (Sb,Bi)$_2$(Se,Te)$_3$ prototypes: it is not charge balanced, which results in bulk metallic states that coexist with
surface Dirac electrons~\cite{Ji_16}. Here, the proposed topological behavior is due to an avoided crossing between valence and conduction bands whose degeneracy is lifted by the
strong spin orbit coupling from the zirconium ions. This intriguing observation raises several questions including whether bulk metallic states can coexist with topologically
protected surface states and, if so, whether the behavior of this composite system is distinct from that of a conventional metal. 




In order to probe the electronic phase space for insights to these issues, we synthesized the chemical analogues (Ti,Zr,Hf)$_2$Te$_2$(P,As), where the strength of the spin-orbit interaction is systematically varied by replacing Ti $\rightarrow$ Zr $\rightarrow$ Hf $\rightarrow$ and P $\rightarrow$ As ~\cite{chen_16_2}. Band structure calculations for the bulk electronic states show precise agreement with the Fermi surfaces determined from quantum oscillations in the magnetic torque and angle resolved photoemission spectroscopy (ARPES). The bulk Fermi surfaces are quasi-two-dimensional and are characterized by relatively high charge carrier mobilities. Calculations predict the presence of composition dependent and topologically protected surface states within their Brillouin zone (BZ).  Angle-resolved photoemission spectroscopy measurements uncover Dirac-like electronic dispersions on the surface of these materials in broad agreement with these predictions. These results allow us to understand how variations in the spin orbit coupling controls the electronic behavior and set the stage to systematically tune the electronic state. 

Importantly, the conversion from topological insulator to a strong topological metal is achieved not through chemical substitution, which would introduce disorder, but rather in stoichiometric materials with nearly perfect crystalline order. This enables the surface states, which have a strong topological character, to realize their ``robust" nature and suggests that the 221s are hosts for potential fundamental discoveries and applications, including: (i) Superlattices of these materials could serve as platforms for topological superconductivity due to the proximity effect, where the necessary metallic component is intrinsic to the system. (ii) Since there are active topological surface states corresponding to a high symmetry point of the Brillouin zone that are close to the Fermi energy, it is possible to study the resulting two dimensional electron gas and its phases. (iii) The interplay between the surface and bulk Fermi sea might produce novel phenomena. (iv) Chemical substitution or intercalation could be used to vary the Fermi energy or the band structure. For instance, the transition from Zr$_2$Te$_2$P to Ti$_2$Te$_2$P is of interest because the former hosts a surface state with strong topological character at the zone center while the latter does not. (v) Any one of these materials or their chemical variants might be useful as components in junction devices. Thus, the 221s represent a deep reservoir for novel physics and have the advantage that their experimentally measured electronic state is fully understood via electronic structure calculations, bringing a design-based methodology within reach.

The paper is organized as follows. In  Sec.\ref{QOE} 
we present the results of our quantum oscillation experiments (QOE). In Sec.\ref{Fermi-Surface}
we give the results for the Fermi surface of these materials
as revealed by our DFT calculations, the QOE and ARPES.
Sec.~\ref{band-structure} we present our results and topology of the
bands as obtained by our DFT calculations and verified by ARPES.
In Sec.~\ref{discussion} we discuss the implications of our experimental and
theoretical findings and in Sec.~\ref{conclusions} we present our
conclusions.


\section{Results}
\subsection{Quantum Oscillation Experiments}
\label{QOE}

\begin{table}[htb]
        \begin{tabular}{l l l l l l l l l l l l}
			\hline
                 & $\alpha$$_{\rm{ZrP}}$ & $\alpha$$_{\rm{HfP}}$ & $\beta$$_{\rm{ZrAs}}$  & $\beta$$_{\rm{ZrP}}$  & $\beta$$_{\rm{HfP}}$ & $\gamma$$_{\rm{HfP}}$ & $\delta$$_{\rm{ZrP}}$  & $\delta$$_{\rm{HfP}}$  \\
            \hline
            $F$(T) & 11 & 107 & 46 & 90 & 149 & 460 & 5800 & 5670\\
            $m^*(m_{\rm{e}}$) & 0.046 & 0.27 & 0.1 & 0.2 & 0.22 & ---  & 1.26 & 1.1\\
			$T_{\rm{D}}$(K) & 19 & 15 & 23 & 11 & 6.5 & ---  & --- & ---\\
			$k_{\rm{F}}$(nm$^{-1})$ & 0.18 & $k_{ab}$ $=$ 0.56 & 0.37 & 0.52 & 0.67 & 1.18 & 4.2 & 4.15\\
			& & $k_{c}$ $=$ 2.08 & & & & & & \\
            $\phi$$_{\rm{B}}$($\pi$) & 1.15 & 0.53 & 1.41 & 1.34 & 1.36 & --- & --- & --- \\
			\hline
        \end{tabular}
        \caption[]{Summary of the properties extracted for the $\alpha$, $\beta$, $\gamma$, and $\delta$ branches of the dHvA signal from Zr$_2$Te$_2$As, Zr$_2$Te$_2$P, and Hf$_2$Te$_2$P. Quantum oscillatory frequencies $F$(T), charge-carrier effective-masses  $m^*(m_{\rm{e}}$), Dingle temperatures $T_{\rm{D}}$(K), and Fermi wave-vectors $k_{\rm{F}}$(nm$^{-1})$ are tabulated.}
        \label{tbl1}
\end{table}

\begin{figure*}
    \begin{center}
        \includegraphics[width= 16 cm ]{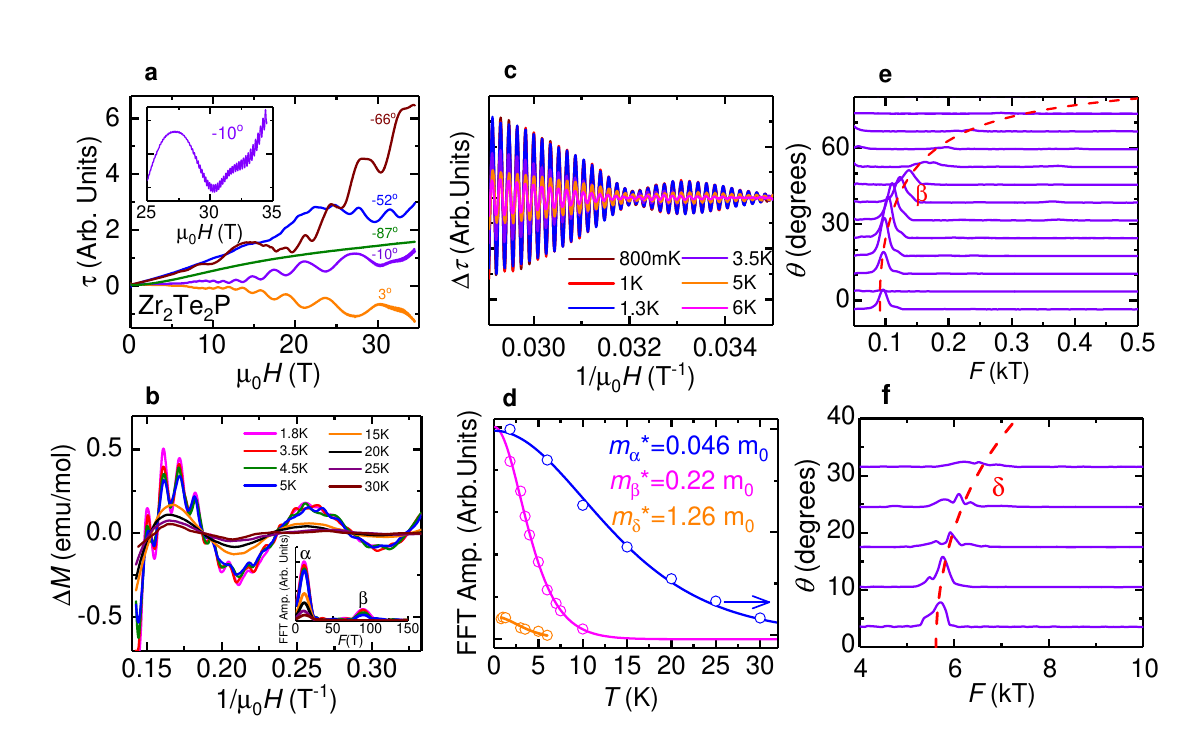}
        \caption{\textbf{Summary of quantum oscillation results for Zr$_2$Te$_2$P.} (a) $\tau$($\mu$$_{0}$$H$) at various angles $\theta$, where low (main panel) and high (inset) frequency de dHvA oscillations are seen. $\theta$ is defined as the angle between the crystallographic $c$-axis and $\mu$$_{0}$$H$, where $\theta$ $=$ 0$^\circ$ and 90$^\circ$ correspond to fields parallel to the $c$- and the $a$-axes, respectively. (b) Taken from Ref.~\cite{chen_16_2}: Background subtracted magnetization $\Delta$$M$ vs. $\mu$$_{0}$$H$ showing the low frequency de Hass-van Alphen (dHvA) oscillations for different temperatures. (inset) Fast Fourier transforms (FFT) for $\Delta$$M$ at various $T$s showing the low frequency $\alpha$ and $\beta$ pockets. (c) Background subtracted $\Delta$$\tau$($\mu$$_{0}$$H$) emphasizing the high frequency de dHvA oscillations. (d) Amplitudes of the low and high frequency peaks observed in the FFT spectra as a function of $T$. Solid lines are fits to the Lifshitz-Kosevich formula from which the effective masses $m^*$ of the charge carriers for different branches are obtained. (e) Fast Fourier transforms of the dHvA signal collected at angles -40$^\circ$ $\leq$ $\theta$ $\leq$ 90$^\circ$ as functions of the cyclotron frequency $F$. The red dashed line is a fit to the low frequency orbit $\beta$, see main text. (f) Fast Fourier transforms of the dHvA signal collected at angles 0$^\circ$ $\leq$ $\theta$ $\leq$ 32$^\circ$ as functions of the cyclotron frequency $F$. The red dashed line is a fit to the high frequency orbit $\delta$, see main text.
        }
		\label{fig:QO}
    \end{center}
\end{figure*}

The Fermi surface topographies and the associated bulk charge carrier properties were probed using torque magnetometry $\tau(\mu_0H)$ measurements (Fig.~\ref{fig:QO} and Supplementary Materials), which reveal de Haas van Alphen (dHvA) oscillatory signals for Zr$_2$Te$_2$P, Hf$_2$Te$_2$P, and Zr$_2$Te$_2$As. Measurements were made at select angles $\theta$ between an external magnetic field $\mu_0 H$ and the crystallographic $c-$axis. $\tau$($\mu_0 H$) was also measured for Ti$_2$Te$_2$P (not shown) but no dHvA signal is present, which is likely due to a lower sample quality or differences in the Fermi surface. The results for Zr$_2$Te$_2$P, Hf$_2$Te$_2$P, and Zr$_2$Te$_2$As are qualitatively similar and in the main text we focus on Zr$_2$Te$_2$P. 

Several families of oscillations labeled $\alpha$, $\beta$, and $\gamma$ are identified for $F$ $<$ 500 T, and close to $\theta = 0^{\circ}$ there is another group of high frequency oscillations ($F$ $\approx$ 5 $-$ 6 kT) labeled $\delta$ for Zr$_2$Te$_2$P and Hf$_2$Te$_2$P. Together, these results reveal two distinct sets of Fermi surface sheets, as summarized in Table~\ref{tbl1}, that are in good agreement with the expectations from electronic structure calculations where $\alpha$, $\beta$, and $\gamma$ are associated with the ``cigar" and ``gear'' hole pockets and $\delta$ with the large electron pockets (see Sec. B and Fig.~\ref{fig:DFTFS} for definitions from electronic structure calculations). The angle dependences of the $\beta$ and $\delta$ frequencies are shown in Fig.~\ref{fig:QO}(e,f) and both are described by the expression $F$$_{\rm{\beta,\delta}}$($\theta$) $=$ $F$(0)/$\cos$$\theta$, as expected for cylindrically shaped Fermi surfaces. Earlier work revealed a cylindrical $\alpha$ branch for Zr$_2$Te$_2$P~\cite{chen_16_2}, but this feature is less prominent in the measurements shown here. Also note that an $\alpha$ orbit has not yet been seen for Zr$_2$Te$_2$As. For Hf$_2$Te$_2$P the $\alpha$ orbit is best described by an ellipsoidal FS and an additional frequency $F$$_{\rm{\gamma}}$ appears at limited angles. The Fermi-wave numbers $k_F$ resulting from these fits are summarized in Table~\ref{tbl1}.


Insight into the nature of the charge carrier quasiparticles is gained by considering the temperature dependence of the dHvA signal amplitude $A$ (Fig.~\ref{fig:QO}d). The $A(T)$ curves are described by the Lifshitz-Kosevich (LK) expression for a Fermi liquid, where fits yield typical bare electron masses $m_\delta$ for the $\delta$ orbit and small effective masses $m_\alpha$ and $m_\beta$ for the small $\alpha$ and $\beta$ orbits, respectively (results for all three compounds are summarized in Table~\ref{tbl1}). This reveals that while the $\alpha$ and $\beta$ pockets have band masses similar to those seen in other topological materials, the large $\delta$ pockets are more conventional in nature. Particularly noteworthy is the very small mass that is observed for the $\alpha$ pocket in Zr$_2$Te$_2$P ($m_\alpha$ $=$ 0.046 $m_e$). 

\subsection{Electronic Fermi Surface} 
We next discuss the calculated Fermi surfaces (FS) determined from density functional theory (DFT) and compare with results from QO and ARPES measurements.
The FS of Zr$_2$Te$_2$P in the rhombohedral unit cell is shown in Fig.~\ref{fig:DFTFS}a, and similar representations for the other three compounds are shown in the Supplemental Materials section (Fig.~S5) . 
The large ``petal''-like electron sheets (blue) are similar in all 4 compounds. In Fig.~\ref{fig:DFTFS}(b) and (c) the projected Fermi surface on the
  $k_x-k_y$ and $k_x-k_z$ planes as measured by ARPES are shown for comparison with the DFT, where there is strong agreement. 
In addition, the quasi-2D Fermi surface structure is confirmed by the measured out-of-plane ARPES Fermi-surface map, shown in Fig.~\ref{fig:DFTFS}(c).
There are also ``hole'' pockets (red) around the $\Gamma$ point which differ between
the 4 materials (Figs.~\ref{fig:DFTFS}(d,e,f,g)). 
 The wavy-features of some of the hole 
pockets are artifacts of the finite k-point mesh and the 
interpolation scheme used in the Xcrysden software\cite{Kokalj}. 
We have shown (see Supplemental Material Fig.~S7) that they can be removed by
using Wannier interpolation but it is computationally costly.
As seen in Figs.~\ref{fig:DFTFS}(d,e,f,g) there is a hole-pocket which can be identified as 
``cigar''-like  and another hole-pocket which can be identified as ``gear''-like. 
The cigar and gear pockets correspond to the two bands at the $\Gamma$ point (labeled as 4 and 3, respectively, in Fig.~\ref{fig:DFTbands}) which intersect the Fermi surface 
for Zr$_2$Te$_2$P, Zr$_2$Te$_2$As, and Hf$_2$Te$_2$P. In the case of Ti$_2$Te$_2$P the lowest of the two bands sinks in the Fermi sea, and the cigar-like pocket disappears.
 For the Hf-based compound this lowest band is also below the Fermi level at the $\Gamma$ point; however, for $\vec k$ near the  high symmetry point $A$ (0,0,1/2) of the hexagonal Brillouin zone boundary, this band moves above the Fermi-level as is the case for Zr$_2$Te$_2$P and Zr$_2$Te$_2$As.

\label{Fermi-Surface}
\begin{figure*}
    \begin{center}
        \includegraphics[width=16 cm]{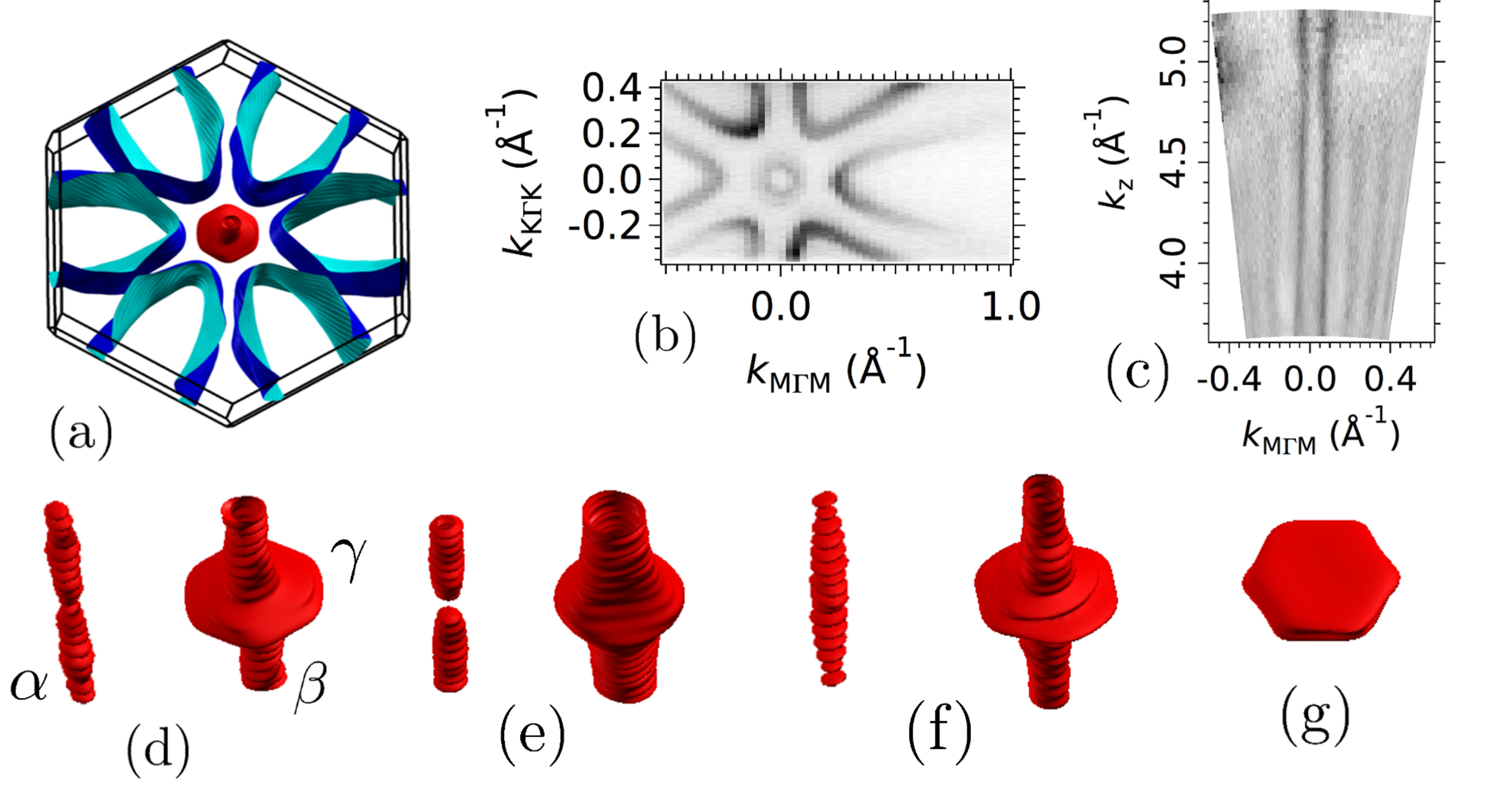}
       \caption{\textbf{Fermi surface sheets calculated from density functional theory (DFT)}:
(a) Calculated full Fermi surface for the Zr$_2$Te$_2$P. 
(b) The projected Fermi surface on the $k_x-k_y$ plane as measured by
ARPES. (c) The projected Fermi surface on the $k_x-k_z$ plane as measured by
ARPES. (d), (e), (f), (g) are respectively the 
 Zr$_2$Te$_2$P,  Hf$_2$Te$_2$P, Zr$_2$Te$_2$As, and Ti$_2$Te$_2$P
hole pockets around the $\Gamma$ point. }
        \label{fig:DFTFS}
    \end{center}
\end{figure*} 
 
The calculated values of the orbits, corresponding to extremal cross-sectional areas of the Fermi surface, viewed along the hexagonal $c-$axis are listed in 
Table~\ref{table2}. The cigar-shaped hole-pocket has one minimum orbit at $\Gamma$ and another one at the $A$ (0,0,1/2) high symmetry point of the hexagonal BZ and a maximum at (0,0,1/4) (shown as $\alpha$ in Fig.~\ref{fig:DFTFS}(d)). The gear-shaped hole-pocket has a maximum at $\Gamma$
(shown as $\gamma$ in Fig.~\ref{fig:DFTFS}(d))  and one minimum orbit at $(0,0,1/2)$ (shown as $\beta$ in Fig.~\ref{fig:DFTFS}(d) which corresponds to the ``gear-shaft''). 
Notice that the Ti-based compound is missing all the orbits which correspond to the cigar shaped hole-pocket. In addition, the Hf-based material is missing the minimal orbit at $k_z=0$. The gear-shaped hole pocket gives rise to a small ($<100$ T) orbit and a medium (several hundred Tesla) orbit. The electron sheets give rise to three different large orbits which correspond to intersections to the hole-pocket at $(0,0,0)$, $(0,0,1/4)$ and $(0,0,1/2)$.

\begin{table}[htb]
\begin{tabular}{|m{75pt}|m{36pt}|m{38pt}|m{35pt}|m{35pt}|} \hline
  FS pocket  & Zr$_2$Te$_2$P  & Zr$_2$Te$_2$As  & Hf$_2$Te$_2$P   & Ti$_2$Te$_2$P \\ \hline
 ``cigar'' ($k_z$=0)            & 26(3)    &13(1)    & 0  &0\\ \hline
 ``cigar'' ($k_z= 1/4$) & 45(1)    & 49(1)   & 49(27) &0 \\ \hline
 ``cigar'' ($k_z= 1/2$) & 26(3) & 13(1) & 49(27) & 0\\ \hline
 ``gear'' ($k_z= 1/2$)               & 60(16)   & 52(9) &172(28) & 0\\ \hline
 ``gear'' ($k_z=0$)    & 792(176)   & 596(192)  &  843(99) & 863(50) \\ \hline
 e-sheet ($k_z =0$) 	& 3805   & 3599  & 4158 & 3641   \\ \hline
 e-sheet ($k_z = 1/4$)     & 7874   & 7226  & 7607  & 6349\\ \hline
 e-sheet ($k_z = 1/2$)     & 6553   & 5790  &  6463  & 7059\\ \hline
\end{tabular}
\caption{Calculated extremal cross-sectional areas for the electron pockets of all four compounds in units of ``Tesla".}
  \label{table2}
\end{table}

The DFT calculations correctly predict the general characteristics
of the Fermi surface sheets as seen by our quantum oscillation experiments.
As shown in Table~\ref{table2}, the calculations predict three very large
electron pockets with frequency in the range of 4000-7000 T and
the quantum oscillation experiments in Table~\ref{tbl1} also reveal frequencies at around 6000 T.
There are several much smaller-size hole orbits originating from the extrema of the
two hole-pocket sheets, which are due to their shape
shown in Fig.~\ref{fig:DFTFS}(d-f). From the size of these frequencies and their angular dependence we can identify the frequencies labeled 
$\alpha$ and $\beta$ in the angular dependence of the QO measurements. 
In both the calculation and the experimental results
the ``cigar''-like hole pocket 
is absent in the case of
the Ti-based compound. This is confirmed by ARPES measurement as shown in Figs.\ref{ARPESKGK}(c,d). 
The band is still present in the Ti-based compound but the pocket
is ``pushed'' below the Fermi-level. In addition, the ``gear'' -like pocket for the Ti-based compounds
loses its ``shaft''.
We note that in going from Zr$_2$Te$_2$P to Zr$_2$Te$_2$As, the frequency
of the small hole pockets is reduced in both the experimental results and
in our computations. We also find that in the case of the Hf-based compound
the ``cigar''-like hole-pocket misses the lowest orbit at $\Gamma$ because
the ``cigar'' thickness at $\Gamma$ seems to vanish.

In the case of the Hf-based compound there is an additional
hole-pocket  (labeled $\gamma$, see Supplemental Material Fig.~S1 (f)) with frequency at around 450 T that is observed experimentally. This
orbit could correspond to the hexagonal-shaped maximum orbit obtained in
all compounds from the ``gear''-like shaped hole-pocket in the
DFT calculations (labeled as $\gamma$ in Fig.~\ref{fig:DFTFS} (d)) with a frequency in the
range of 500-900 T. Its precise value is very sensitive to very small
adjustments of the Fermi surface and the discrepancy from the
experimental value should not be taken seriously. Furthermore, its absence from
the other compounds may be due to a lack of the required
experimental sensitivity, as even in the Hf-based compound it is difficult to detect.

\subsection{Electronic Band Structure} 
\label{band-structure}
The calculated band-structures, with and without spin-orbit coupling (SOC), along with their orbital character for Zr$_2$Te$_2$P are presented in Figs.~\ref{fig:DFTbands}(a),(b).
 A first indication of the presence of non-trivial topology is seen by comparing the band structure with and without SOC. 
Between bands 1 and 2, which are of Zr $d$ and Te $p$ orbital character, at the time-reversal invariant momentum (TRIM) \emph{Z}, which projects onto the $\bar{\Gamma}$-point of the hexagonal BZ in our slab calculation and in the notation of our ARPES measurements,  there is a band inversion similar to what is seen for Bi$_2$Se$_3$, although in the latter case the orbitals involved in the band inversion are of $p$ character, which form bands of different parity at the {\bf $\Gamma$} point \cite{Zhang2009}. 
In addition,  at the TRIM $L$ 
(which projects onto the $\bar{M}$-point of the hexagonal BZ), 
there is parity inversion between bands 2 and 3. 
Notice that these two bands are made of Te $p$ orbitals as in the Bi$_2$Se$_3$ case.  Similar features are found for the Zr$_2$Te$_2$As and  Hf$_2$Te$_2$P compound. In the case of the Ti$_2$Te$_2$P compound, while there is parity inversion at the 
${\bar M}$ point, there is no inversion
at  the $\bar{\Gamma}$-point.
As a result all our compounds have a topological surface state between bands 2 and 3 at the  $\bar{M}$-point and at the $\bar{\Gamma}$ all other compounds have a topological surface state between the aforementioned two bands with the exception of the Ti-based one. Figs.~\ref{fig:DFTbands}(c),(d) present the bulk bands (shown as gray ribbons) 
with bands obtained from a slab calculation (shown as blue lines) 
depicting the presence of a
Dirac-like surface state at the $\bar{\Gamma}$-point in Zr$_2$Te$_2$P in panel (c)
and the absence of such a surface state for Ti$_2$Te$_2$P at the $\bar{\Gamma}$-point
({see Supplemental Material Figs.~S8 and S9 for the surface states in Hf$_2$Te$_2$P.)
The Ti-based compound does not have strong enough SOC to cause band inversion at the $\bar{\Gamma}$ point; at the ${\bar M}$-point, however, the band character in all compounds, including the Ti-based one, is Te $p$-type for which the SOC is strong enough to
cause band inversion. 
 \begin{figure}[htb]
    \begin{center}
 \includegraphics[width=1\linewidth]{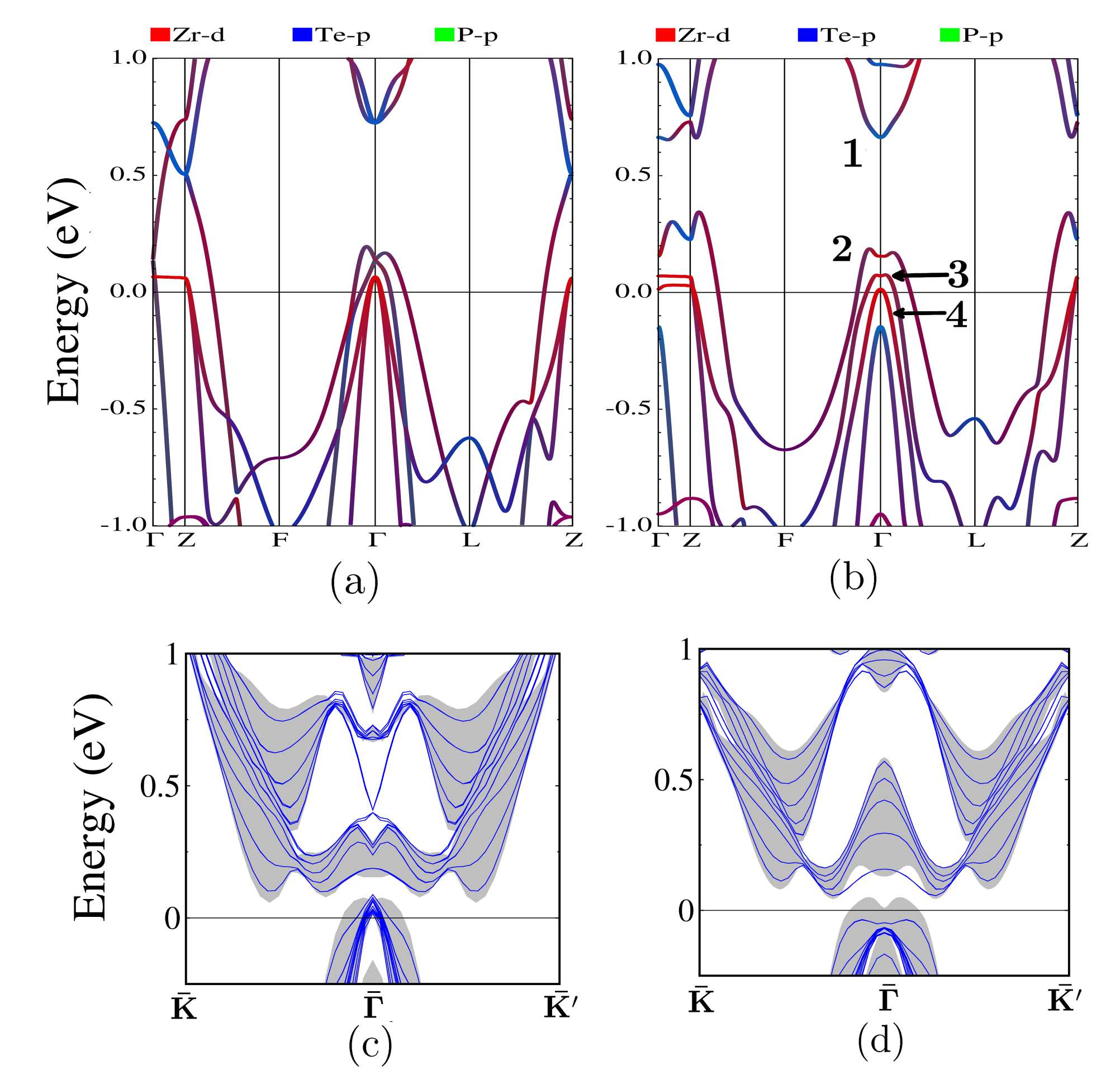}
	\caption{
	\textbf{Band Structure of Zr$_2$Te$_2$P in the rhombohedral unit cell along the TRIMs} (a) 
	 without the inclusion of the spin-orbit coupling (SOC) and (b) with the inclusion of SOC. 
	The colors used indicate the orbital character of the bands: red, blue and green colors
	corresponds to Zr \emph{d}-, Te $p-$ and P $p-$orbitals, respectively.
	The superposition of bulk bands (shown as Gray ribbons) with 
bands obtained from a slab of 5 quintuple layers (shown as blue lines) 
         depicting the presence of a
	 Dirac-like surface state at the $\bar{\Gamma}$-point is shown for Zr$_2$Te$_2$P in panel (c). Panel
	(d) shows the absence of such a surface state for Ti$_2$Te$_2$P at the $\bar{\Gamma}$-point.}

	\label{fig:DFTbands}
    \end{center}
\end{figure}

In order to investigate whether or not the Dirac surface states seen in ARPES and slab calculation have strong topological character, 
we performed parity analysis
of the bulk bands following Ref.~\onlinecite{Fu_TI_Inversion_Symmetry}, which is simplified because of the presence of inversion symmetry in the 221 structure.  
We first focus on the surface state around the $\bar{\Gamma}$ point that is present in Zr$_2$Te$_2$P and Hf$_2$Te$_2$P but is absent in Ti$_2$Te$_2$P (see Fig.~\ref{fig:DFTbands} (c) and (d)), as was done previously in Ref.~\onlinecite{Ji_16}. 
We find the parity product of all the occupied bands below band 1, where we neglect the deeply situated bands that are well separated from the higher occupied bands. For Zr$_2$Te$_2$P and Hf$_2$Te$_2$P, 
the parity product at the  TRIMs around the $\Gamma$-, \emph{F}-, \emph{L}- and \emph{Z}-points within the BZ \cite{Zhang2009} are respectively positive, positive, positive and negative 
whereas the parity product for all the TRIMs in Ti$_2$Te$_2$P is positive. 
This gives a Z$_2$ topological index $\nu_0=1$ for Zr$_2$Te$_2$P and Hf$_2$Te$_2$P, whereas $\nu_0=0$ for Ti$_2$Te$_2$P. This implies that there are an odd number of surface states between bands 1 and 2 for Zr$_2$Te$_2$P and Hf$_2$Te$_2$P and that they are robust to external perturbations \cite{Fu_TI_Inversion_Symmetry}, 
whereas there are even (or zero) number of surface states for Ti$_2$Te$_2$P which are not robust. 
Note that the TRIM \emph{Z}, where parity inversion is seen, projects onto the $\bar{\Gamma}-$point in the hexagonal Brillouin zone.
We refer the reader to Refs.~\onlinecite{Zhang2009} and ~\onlinecite{Ji_16}  for an illustration of how the rhombohedral BZ projects onto the 2D hexagonal BZ.

A similar analysis was performed to investigate the topology of the surface state seen between bands 2 and 3, i.e., those leading to the ``petals'' and the ``gear''-like FSs respectively, at the $\bar{M}-$point. 
From this analysis we find $\nu_0 = 1$ for Zr$_2$Te$_2$P and Ti$_2$Te$_2$P, whereas $\nu_0 = 0$ for Hf$_2$Te$_2$P. For Zr$_2$Te$_2$P and Ti$_2$Te$_2$P, the parity product  is negative only at the TRIM \emph{L} which projects onto $\bar{M}$ point of the hexagonal BZ, whereas for Hf$_2$Te$_2$P the parity product is negative at the two TRIMs \emph{Z} and \emph{L}. 
This makes the overall product positive and results in a $Z_2$ index of 0. Thus, we expect that for Zr$_2$Te$_2$P and Ti$_2$Te$_2$P, 
there are  odd numbers of Dirac-like states (three at the \emph{L-}points which projects onto the $\bar{M}-$point of the hexagonal Brillouin zone of the slab)  and for Hf$_2$Te$_2$P, there are even number of Dirac-like states (one at the $\bar{\Gamma}$ point and three at the $\bar{M}-$point) between the electron band and the ``gear" band. 
Therefore, for Hf$_2$Te$_2$P, the parity analysis shows that the surface Dirac-band at the $\bar{M}$-point  does not have strong topological character because the parity inversion happens at both the \emph{Z}-  and \emph{L}- point as stated above.

\begin{figure}
    \begin{center}
        \includegraphics[width= 8.6 cm]{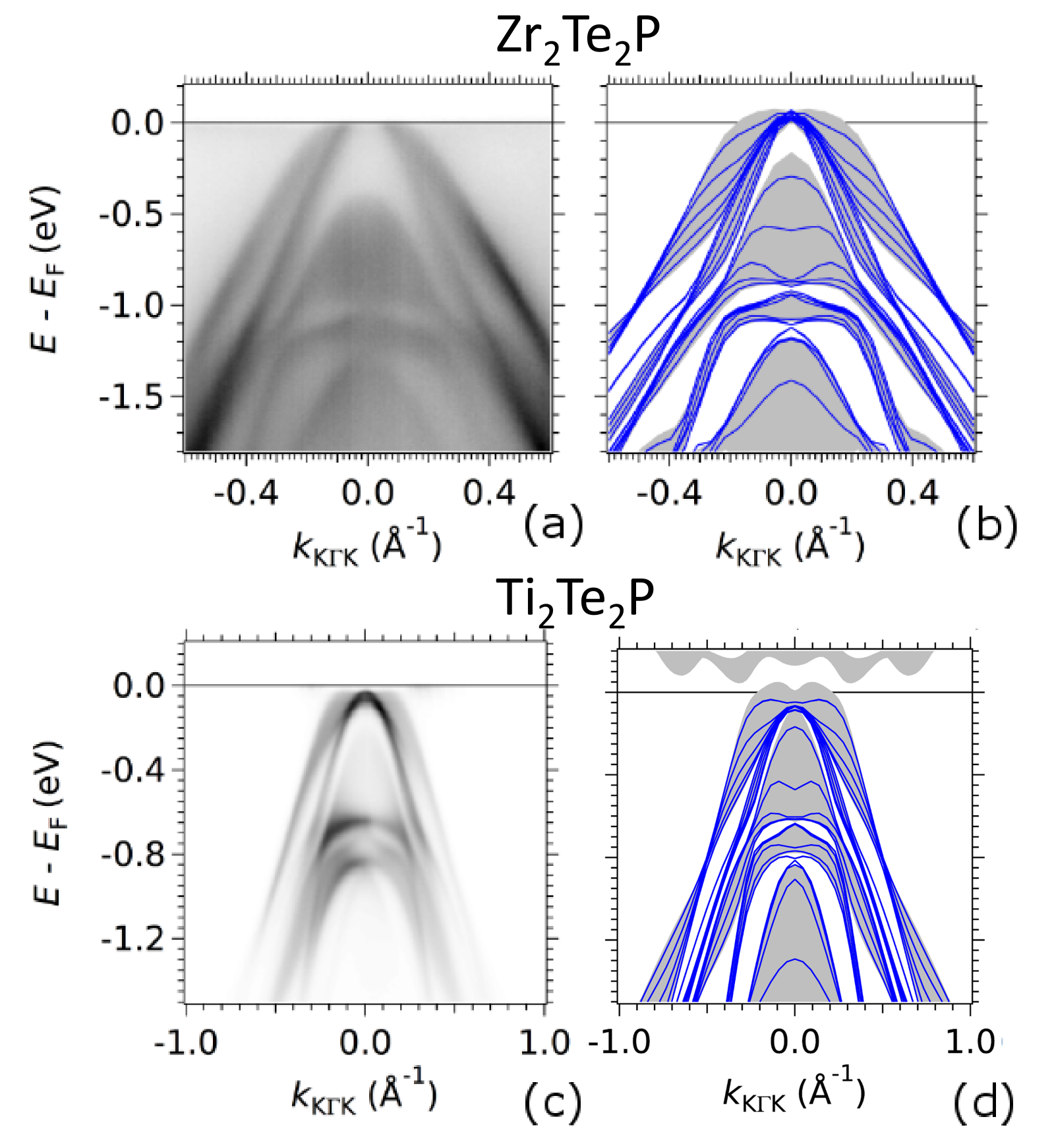}
       \caption{\textbf{Comparison between ARPES measurements and band structure calculations for cuts along the $K$ - $\Gamma$ - $K$ direction.}  (a) ARPES determined band dispersion and (b) calculated DFT bands for Zr$_2$Te$_2$P, respectively. (c) ARPES and (d) DFT for Ti$_2$Te$_2$P, respectively. Fermi energy $E_{\rm{F}}$ is indicated by the horizontal black line. Blue solid lines are bands obtained from a 5-layer slab calculation and therefore represent surface states and grey ribbons represent the bulk bands.
       }
        \label{ARPESKGK}
    \end{center}
\end{figure}

\begin{figure}
    \begin{center}
        \includegraphics[width= 8.6 cm]{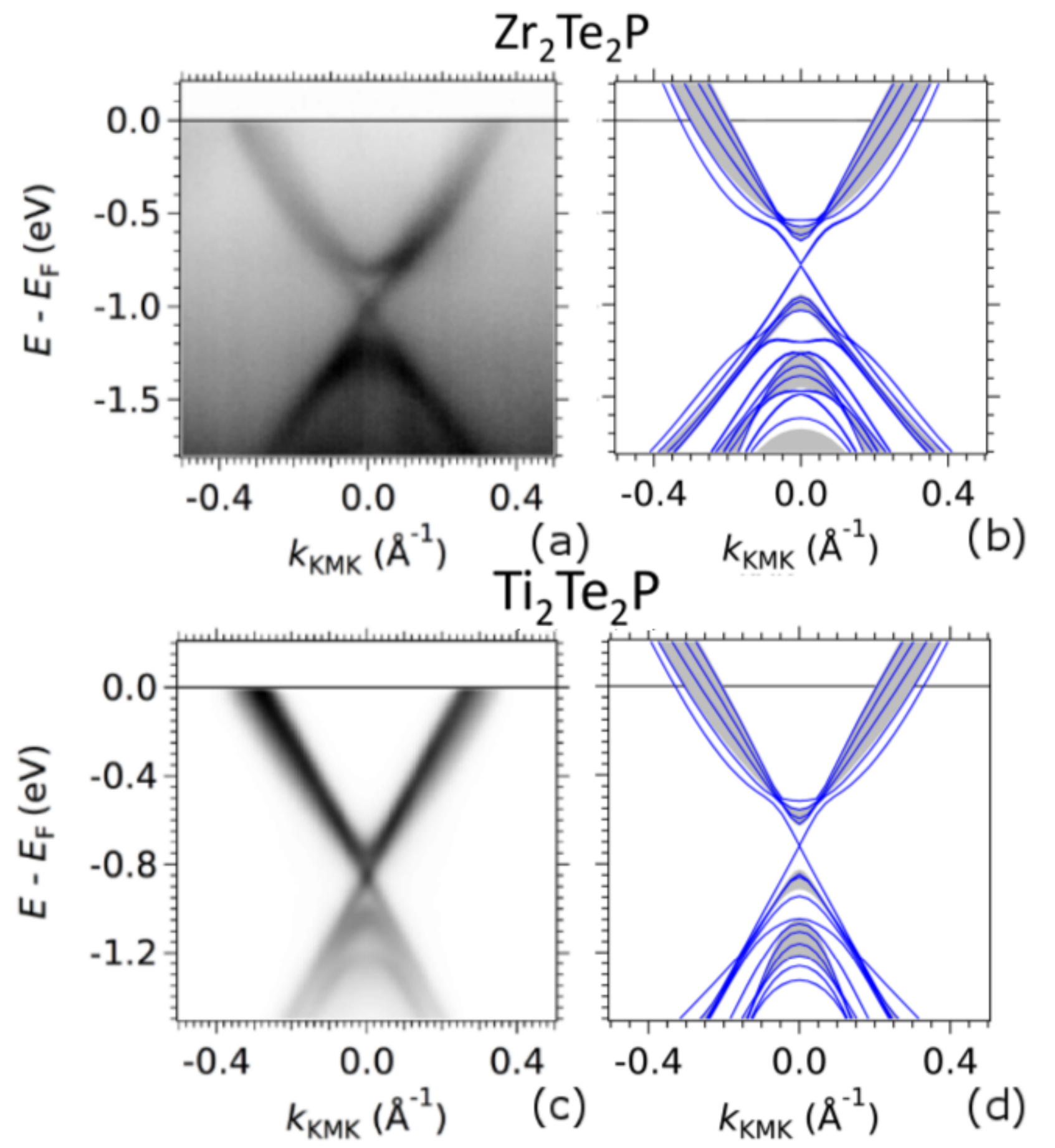}
       \caption{\textbf{Comparison between ARPES measurements and band structure calculations for cuts along the $K$ - $M$ - $K$ direction.} (a) ARPES determined band dispersion and (b) calculated DFT bands for Zr$_2$Te$_2$P, respectively.  (c) ARPES and (d) DFT for Ti$_2$Te$_2$P, respectively.  Fermi energy $E_{\rm{F}}$ is indicated by horizontal line. Solid blue lines depict bands obtained from a 5-layer slab calculation and represent surface states and grey ribbons represent the bulk bands.}
        \label{ARPESKMK}
    \end{center}
\end{figure}
Fig.~\ref{ARPESKGK} shows a comparison between the surface band structure measured by ARPES and the one calculated by DFT for both Zr$_2$Te$_2$P (panels a and b) and Ti$_2$Te$_2$P (panels c and d) along the $\bar{K}$ - $\bar{\Gamma}$ - $\bar{K}$ direction. Hole-like states are observed in both cases, which agree well with the calculated ones.  Likewise, Fig.~\ref{ARPESKMK} compares the measured (ARPES) and calculated (DFT) surface band structure for the same two compounds along the $\bar{K}$ - $\bar{M}$ - $\bar{K}$ direction. In both cases, a Dirac cone, with its vertex located at around 1 eV below $E_F$ for Zr$_2$Te$_2$P (0.82 eV below $E_F$ for Ti$_2$Te$_2$P), is observed in the ARPES data (panels a and c). The same Dirac cone is predicted by DFT (panels b and d). 

As seen in Figs.~\ref{ARPESKGK} and ~\ref{ARPESKMK}, the ARPES and DFT band structure is in good agreement, especially close to the Fermi level.
Moreover, the presence or absence of certain hole pockets, such as the ``cigar'' or ``gear''-like pockets, moving from one material studied to other, can be verified by examining the intersection of the
ARPES bands with Fermi surface as predicted by the DFT calculations and seen in our QO experiments. We highlight the absence of the ``cigar''-like pocket for Ti$_2$Te$_2$P as correctly predicted by DFT and confirmed by ARPES.

For the bands away from the Fermi level, the DFT calculations have to be up-shifted with respect to the ARPES ones by 100 - 200 meV with this shift being not rigid. 
For instance, between the $\bar{K}$ - $\bar{\Gamma}$ - $\bar{K}$ and $\bar{K}$ - $\bar{M}$ - $\bar{K}$ directions the amount of shift is compound dependent. We however note that the energy position of the bands crossing $E_F$ is almost unchanged in ARPES and DFT. 
Surface band bending effects -typical in tetradymite topological insulators \cite{bianchi,frantzeskakis}- may be at the origin of such energy shifts of the experimental bands. 
In any case, such a shift is a purely extrinsic effect that does not change our main analysis and conclusions. The results for Hf$_2$Te$_2$P are summarized in the Supplemental Material, where the measured Fermi surfaces for all samples are also shown. For all three compounds there is a good agreement between the calculated and the measured bulk bands, 
while the out-of-plane Fermi surface of Zr$_2$Te$_2$P confirms the two-dimensional character of the bands (see Fig.~\ref{fig:DFTFS} (c)) as suggested by our QO measurements. We therefore conclude that the Fermi level found from the self consistent calculation of the Kohn-Sham Hamiltonian gives the best agreement between theory and experiment.

\vskip 0.3 in

\section{Discussion}
\label{discussion}

Our results comprehensively characterize the electronic states of the layered tetradymite metals (Ti,Zr,Hf)$_2$Te$_2$(As,P) and the impact of changing the spin-orbit coupling through stoichiometric chemical variations. In the bulk, there are both hole and electron pockets which all have close to cylindrical topography, as revealed by electronic structure calculations, the de Haas-van Alphen effect, and ARPES measurements. The hole pockets are small in volume with their associated charge carriers characterized by light cyclotron masses. The electron
sheets are larger with effective masses that are similar to the bare electron mass. Intriguingly, the slab calculations predict linearly dispersing surface states of strong topological character (as shown from parity analysis) above the Fermi energy at the $\bar{\Gamma}$ point for the Hf and Zr versions and below the Fermi energy at the $\bar{M}$ point for all chemical variants, which are indeed observed by ARPES measurements. Therefore, the 221 materials are shown to be an attractive family of materials in which to develop novel phenomena that exploit both the bulk and surface states, where electronic structure calculations provide quantitative guidance for designing specific band structures.

For instance, for the Zr- and Hf-based compounds, the slab calculations and bulk parity analysis predict unoccupied linearly dispersing 
surface states at around 250 meV above the Fermi energy along the $\bar{K}$ - $\bar{\Gamma}$ - $\bar{K}$ direction, similar to what is seen for the binary tetradymite topological insulators~\cite{Xu2014,Chen178}. In order to experimentally observe these states it will be useful to electronically dope these systems to fill the empty bands, as attempted earlier \emph{via} Cu intercalation~\cite{Ji_16}. Importantly, these states are predicted to be absent in the Ti-based compound by virtue of weaker spin-orbit coupling. This provides a natural way to compare behaviors between a system that is expected to exhibit behavior related to a Dirac point and one that is not. There are several ways to accomplish charge doping, including chemical substitution and intercalation, but a particularly attractive route would be to introduce a non-stoichiometric fraction on the pnictogen site, which would fill the bands of the surface state around the $\bar{\Gamma}$-point. This would allow us to investigate if the topological surface states are robust against translational symmetry-breaking chemical disorder. 

It is also important to investigate if it is possible to induce unconventional, topological superconductivity in these compounds. Again, there are several possible routes but an attractive example is to do it through the proximity effect~\cite{proximity}, by pairing carriers on the surface bands around the $\bar{M}$-points. If both the superconductivity and the robustness of the surface states with respect to disorder was confirmed, one could explore additional superconducting gap symmetries, by inducing superconductivity in the aforementioned alloys containing filled surface state bands around both the $\bar{\Gamma}$- and the $\bar{M}$-points. Notice that it might be possible to induce bulk superconductivity just by applying hydrostatic pressure, chemical substitution \cite{doping} or even electrostatic doping in exfoliated specimens~\cite{electrostatic}. In other words, these compounds offer an avenue to induce novel and perhaps topological superconducting states possibly involving the electronic structure of the bulk and of their surface states.

\vskip 0.5 in

\section{Summary}

\label{conclusions}
We have studied a family of strongly topological metals with the chemical
formula $T_2$$Ch_2$$Pn$ ($T$ $=$ Ti, Zr, Hf) which are characterized by nearly perfect 
crystalline order. The electronic band structures and Fermi surfaces for these materials are reported based on electronic structure calculations, quantum oscillations in torque magnetometry, and angle resolved photoemission spectroscopy. The bulk Fermi surface consists of hole and electron pockets which are all nearly two dimensional. The electron sheets are conventional in nature, while the hole pockets have small cyclotron masses. Intriguingly, the slab electronic structure calculations predict topologically protected Dirac-like surface states at two distinct locations within their Brillouin zone. Of these states, those located below the Fermi level are observed by our angle resolved photoemission spectroscopy measurements which are in good agreement with DFT calculations.

These materials provide a platform on which to study surface states of strong topological character in coexistence with an {\it intrinsic} 
bulk metal.  The fact that the metallic character of our crystalline samples is
intrinsic implies that the strong topological surface states present in these crystals, might be long-lived. This could be utilized in several potential 
applications, including quantum computation.
We furthermore find that the strongly topological surface states that are associated with the high symmetry point ${\bar {M}}$
 are not too distant from the Fermi energy. This potentially motivates fundamental 
studies of the low-temperature phases of this two-dimensional (2D) topological interacting electron gas (EG) which 
can be caused by electronic correlations, as in the usual 2DEG. Furthermore, the coexistense of the topologically protected surface
states and the intrinsic metallic component seem to offer a number of opportunities to induce and study unconventional and possibly topological superconductivity following several routes.
For example, superlattices based on these materials might serve as platforms for realizing topological superconductivity due to 
the proximity effect\cite{proximity} because the needed metallic component for the mechanism to work is adjacent to the topological subsystem. Another possibility is to look for topological superconductivity in granular states of this family of materials where the surface contribution is significant through the Josephson-junction effect.
Lastly we found that Zr$_2$Te$_2$P hosts a surface state at the zone center of strong topological character, while Ti$_2$Te$_2$P does not.
Hence, it may be interesting to pursue future studies of this transition from Zr$_2$Te$_2$P to Ti$_2$Te$_2$P by means of atomic substitution.

\section{Acknowledgements}
This work was performed at the National High Magnetic Field Laboratory (NHMFL), which is supported by National Science Foundation Cooperative Agreement No. DMR-1157490, the State of Florida and the DOE. A portion of this work was supported by the  NHMFL User Collaboration Grant Program (UCGP). L.~B. is supported by DOE-BES through award DE-SC0002613. The work at KEK-PF was performed under the approval of the Program Advisory Committee (proposals 2016G621 and 2015S2-005) at the Institute of Materials Structure Science, KEK.


\newpage
\title{Supplemental Materials: "Converting topological insulators into topological metals within the tetradymite family"}

\textbf{Experimental and Calculation Methods}

\textbf{Single crystal growth} Single crystals of $M$$_2$Te$_2$$X$ ($M$ = Ti, Zr, Hf, $X$ = P, As) were grown using the chemical vapor transport method.~\cite{chen_16_2} Stoichiometric polycrystalline precursor material was prepared by reacting raw elements at 1000$^\circ$C for 24 hours. The resulting powders were subsequently sealed under vacuum with 3 mg/cm$^3$ of iodine in quartz tubes. The tubes were placed in a resistive tube furnace with a temperature gradient with 800$^\circ$C (source) and 900$^\circ$C (drain) for 21 days. Exfoliable single crystal specimens with a hexagonal plate shape and dimensions 3$\times$3$\times$1 mm formed in the cold zone of the ampoule. The crystallographic $c$-axis is perpendicular to the plate face.

\textbf{Torque magnetometry and background subtraction}
The magnetic torque $\tau$, which measures magnetic anisotropy under an applied magnetic field $\mu_0$$H$, was measured using a piezoelectric cantilever (SEIKO-PRC400) in 35T resistive magnets at the National High Magnetic Field Laboratory in Tallahassee Florida (Fig.~\ref{fig:angle}). For these measurements, the tilted magnetic field $\mu$$_0$$H$ was confined to the $a$-$c$ plane and $\theta$ was the angle between the $c$-axis and $\mu$$_0$$H$ (Fig.~\ref{fig:background}c). For simplicity, the axes were chosen so that $z$$\parallel$$c$ and  $y$$\parallel$$a$. The torque only occurs in the $x$ direction: $\tau$ = $m$$\times$$\mu_0$$H$ = 1/2$V$$\Delta$$\chi$$\mu_0$$^2$$H$$^2$$\sin$(2$\theta$), where $\chi$ is the magnetic susceptibility, $\Delta$$\chi$ = $\chi$$_a$ - $\chi$$_c$, and $V$ is the volume of the sample. $\tau$($\mu_0$$H$) is composed of a monotonically increasing background and superimposed oscillations due to the de Haas van Alphen (dHvA) effect. The dHvA oscillations are isolated by subtracting a smoothed (moving window average) data curve (dotted line in Fig.~\ref{fig:background} inset) from the $\tau$($\mu_0$$H$) curve. An example background subtracted curve is shown in Fig.~\ref{fig:background}b, where the sign can be identified with the anisotropy $\Delta$$\chi$$_b$ = $\chi$$_c$ - $\chi$$_a$ of the paramagnetic susceptibility.~\cite{chen_16_2} In Fig.~\ref{fig:Mcompare}, we compare $\tau$($\mu_0$$H$) with magnetization $M(H)$ data collected in a Magnetic Property Measurement System (MPMS). The phases of the dHvA oscillations shows a very good agreement between the two measurements. In order to extract the frequency spectrum for each specimen, the oscillations on $\tau$($H$) were fast Fourier transformed (FFT).     


The oscillations are described by the expression $\Delta M$ $\propto$ -$H$$^{1/2}$$R_T$$R_D$$\sin$[{2$\pi$($\frac{F}{H}$-$\gamma$+$\delta$)]~\cite{shoenberg}. Here, $R_T$ $=$ $\alpha$$T$$m^*$/$H$$\sinh$($\alpha$$T$$m^*$/$H$) is the thermal damping factor, $R_D$ $=$ $\exp$(-$\alpha$$T_{\rm{D}}$$m^*$/$H$) is the Dingle damping factor, $m^*$ is the charge carrier effective mass and $\alpha$ $=$ 2$\pi^2$$k_B$$m_e$/$e$$\hbar$ $\approx$ 14.69 T/K. The phase factor $\gamma$ acquires values of 0 and 1/2 for two- and three-dimensional Fermi surfaces, respectively . The Dingle reduction factor $R_{\rm_D}$, describes the broadening of the Landau levels due to scattering.~\cite{shoenberg}. The Dingle temperature is a useful parameter $T_{\rm{D}}$=$\hbar$/2$\pi$$k_B$$\tau$, where $\tau$ is the relaxation time averaged over one cyclotron orbit. By plotting of ln($\tau$/(($\mu_0$H)$^{3/2}$R$\rm_T$) vs 1/$\mu_0$H the Dingle temperature can be extracted by linear regression. A summary of $T_{\rm{D}}$s are given in Table II of the Main Text, showing a variation of electronic scattering in different compounds which reveals differing sample qualities. In particular, we note that the high Dingle temperature for Zr$_2$Te$_2$As indicates a larger density of defects in these samples that might account for the absence of the $\alpha$ and $\delta$ pockets in the quantum oscillation measurements. This is consistent with the smaller residual resistivity ratio of Zr$_2$Te$_2$As when compared to either Zr$_2$Te$_2$P or Hf$_2$Te$_2$P, which is an additional indication of enhanced scattering due to a larger density of defects.





\textbf{Electronic Structure Calculations using Density Functional Theory (DFT)}
Electronic structure calculations were performed using the experimentally determined crystal structures using the Vienna \emph{ab-initio} simulation package~\cite{VASP1,VASP2,VASP3,VASP4}(VASP) and the Quantum Espresso (QE) \cite{QE} implementations of the Density Functional Theory (DFT) within the generalized  gradient approximation (GGA). The Perdew-Burke-Ernzerhof (PBE) exchange correlation  functional~\cite{PBE} and the projected augmented wave (PAW) methodology~\cite{Blochl} was used to describe the core electrons in VASP whereas Optimized Norm-Conserving Vanderbilt Pseudopotentials~\cite{ONCVPP} were used in QE.
The band structure calculations were also verified using the all electron WIEN2k~\cite{wien2k} implementation. 
Spin-orbit coupling (SOC) was included in all of our calculations.

We also employed the Wannier90~\cite{Wannier90} software to calculate maximally
localized Wannier functions (MLWFs) for Zr$_2$Te$_2$P from the Bloch states calculated using QE. 
The Wannierization method enabled us to perform high resolution calculations in
k-space as evidenced from the high resolution Fermi
surface obtained by using a $k-$point mesh of $100\times 100\times 100$ shown
in Fig.~\ref{fig:WannierHole}.
Notice that the wavy-like features seen in the hole pockets are no longer present after Wannierization.
This suggests that these features are artifacts of the finite k-point mesh.
For the Wannierization procedure, 38 MLWFs were constructed using the method of disentanglement\cite{disentanglement} since the valence bands are not separated from the conduction bands by a gap.
All Zr \emph{d}, Te \emph{p} and P \emph{p} states were used in this procedure.

In Figs.~\ref{fig:DFTFS} and \ref{fig:bands}, we present, respectively, the Fermi surfaces and the band structures of all the four crystals 
calculated in the rhombohedral unit-cell.
In Fig.~\ref{fig:HTPARPESDFT},the comparison between the ARPES and DFT  band dispersion is presented for Hf$_2$Te$_2$P.
We also present the DFT calculated band dispersion above the Fermi level ($E_F$) for Hf$_2$Te$_2$P in Fig.~\ref{fig:HTPDFT} which shows the presence of the Dirac-like 
surface state at around 300 meV above the $E_F$ at the $\bar{\Gamma}-$ point.


\textbf{Angle Resolved Photoemission Spectroscopy (ARPES)}
Single crystals of $M$$_2$Te$_2$$X$ were oriented by means of Laue diffraction before mounting on the ARPES sample holders. ARPES experiments were performed at the CASSIOPEE beamline of Synchrotron SOLEIL and at the beamline 2A of the KEK-Photon Factory (KEK-PF) using hemispherical electron analyzers with vertical and horizontal slits, respectively. Pristine sample surfaces were generated by \textit{in situ} cleavage at pressures of the order of 5$\times$10$^{\textmd{-}11}$ mbar. The sample temperature during cleavage and measurements was 5 K (SOLEIL) or 20 K (KEK-PF), without observing any $T$-dependence between these two temperature values. The typical angular and energy resolutions were 0.25$^{\circ}$ and 15 meV, while the mean diameter of the incident photon beam was of the order of 50 $\mu$m (SOLEIL) and 100 $\mu$m (KEK-PF). We used variable energy and polarization of the incident photons. Data shown in Figs. 6 and 7 were acquired with photon energies of 50 eV (Ti$_2$Te$_2$P) and 120 eV (Zr$_2$Te$_2$P), while the polarization was set to linear horizontal. A systematic variation of the photon energy revealed no changes in the energy-momentum dispersion, a feature that is characteristic of 2D-like band structures. During the time window of our measurements the pressure was in the range of 10$^{\textmd{-}11}$ mbar and no surface degradation was observed. \\ 

\begin{figure*}
\renewcommand{\thefigure}{S\arabic{figure}}
    \begin{center}
        \includegraphics[width= 1\linewidth]{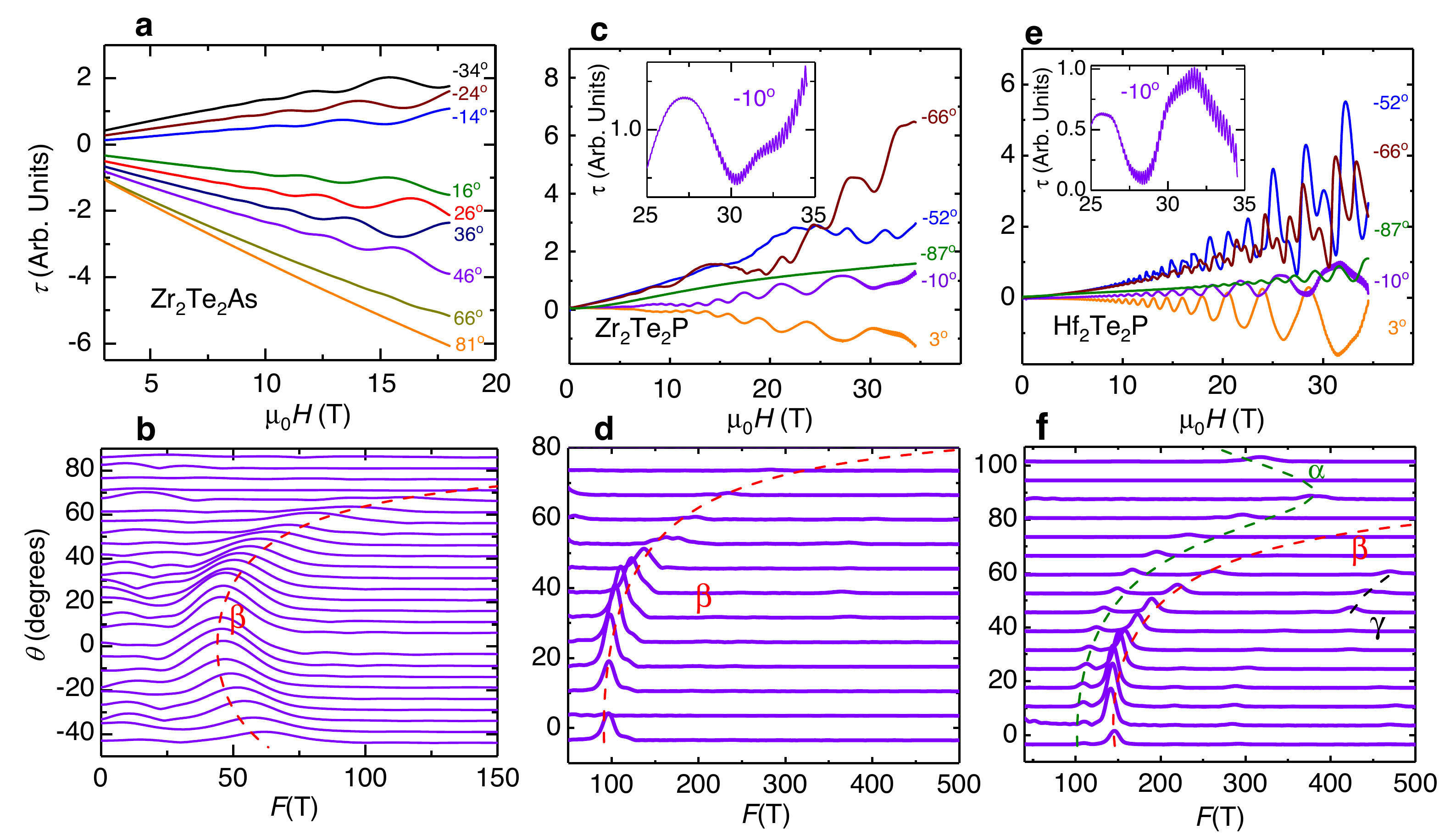}
        \caption{\textbf{Magnetic torque $\tau$ as function of field.}
        Magnetic torque $\tau$ for: (a) Zr$_2$Te$_2$As (c) Zr$_2$Te$_2$P and (e) Hf$_2$Te$_2$P respectively, as a function of magnetic field $\mu$$_{0}$$H$
        for various angles $\theta$. Here, $\theta$ is the angle between the crystallographic $c$-axis and the applied field $\mu$$_{0}$$H$,
        where $\theta$ $=$ 0$^\circ$ and 90$^\circ$ correspond to fields parallel to the $c$- and the $a$-axes, respectively.
        Low frequency de dHvA oscillations are seen for all three compounds. High frequency dHvA-oscillations are highlighted in
        the insets within panels (c) and (e). (b), (d), (f) Fast Fourier transforms of the respective dHvA signals collected at angles -40$^\circ$ $\leq$ $\theta$ $\leq$ 90$^\circ$
        as functions of the cyclotron frequency $F$. Red dashed lines in (b), (d), and (f) depict the low frequency orbits $\alpha$, $\beta$, and $\gamma$
        determined by the electronic band-structure calculations. Lines depicting the $\alpha$ and $\beta$ orbits correspond to fits to the experimental data, see main text.}
        \label{fig:angle}
    \end{center}
\end{figure*}

 \begin{figure}[!tht]
 \renewcommand{\thefigure}{S\arabic{figure}}
    \begin{center}
 \includegraphics[width=1\linewidth]{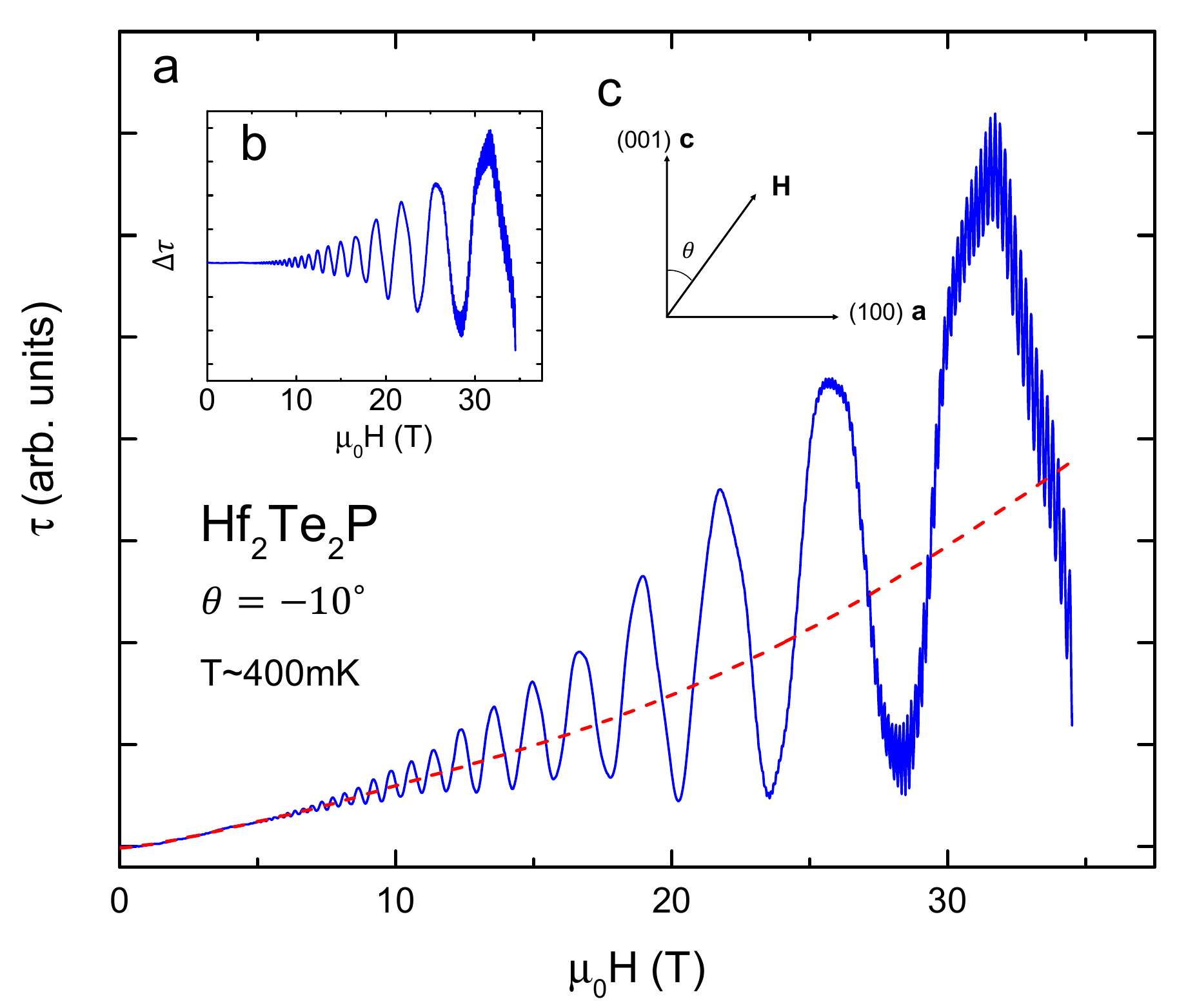}
	\caption{\textbf{Background subtraction to isolate de Haas van Alphen (dHvA) oscillations.}
	 (a) An example data set of the torque $\tau$ versus magnetic field $\mu_0$$H$ for Hf$_2$Te$_2$P. The red dotted curve is the anisotropy paramagnetic background acquired by smoothing the data curve. (b) The oscillatory component of the torque $\Delta$$\tau$ acquired after the background subtraction. (c) Schematic drawing of the orientation of the crystalline axes with respect to $\mu_0$$H$. }
	\label{fig:background}
    \end{center}
\end{figure}

 \begin{figure}[!tht]
 \renewcommand{\thefigure}{S\arabic{figure}}
    \begin{center}
 \includegraphics[width=1\linewidth]{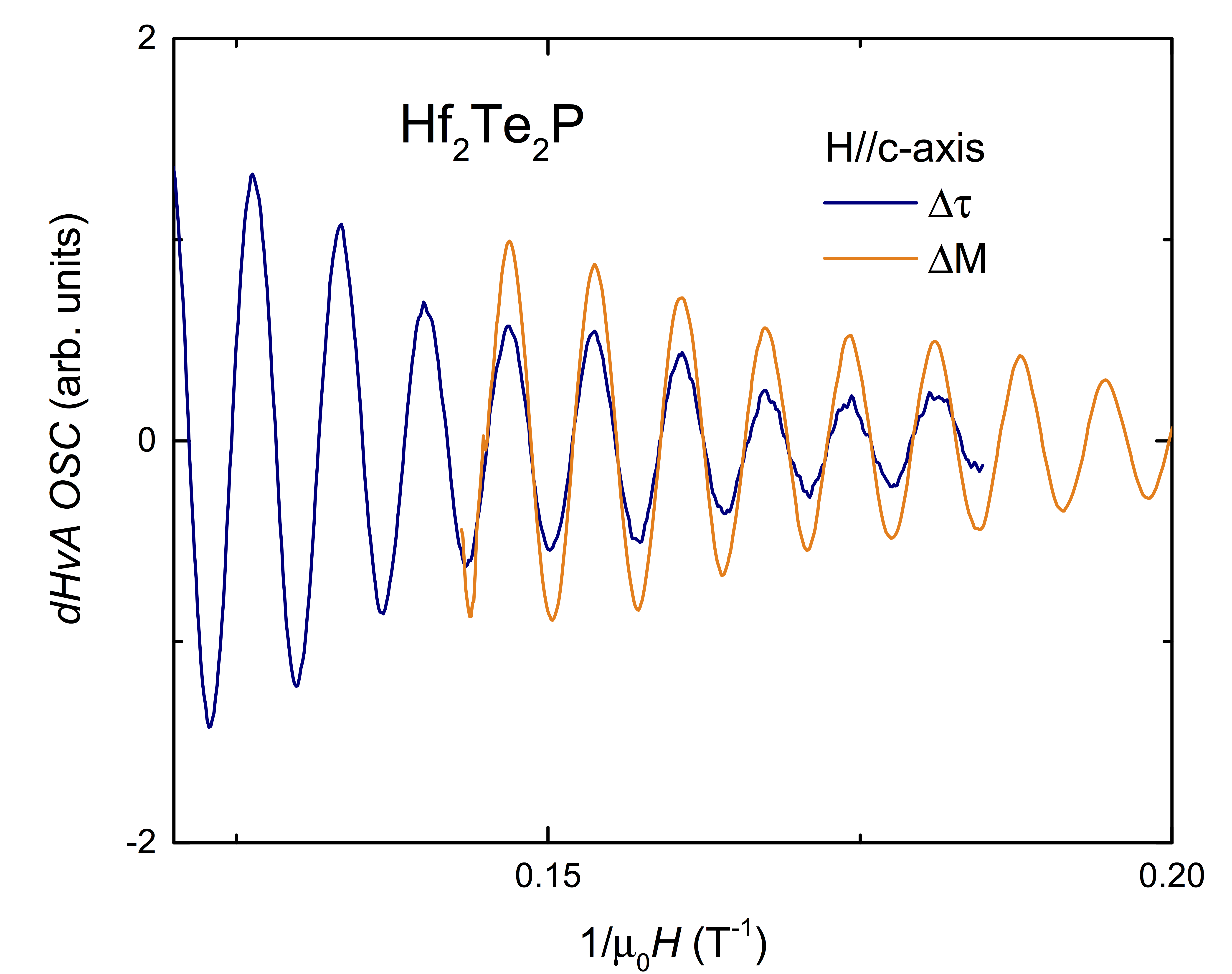}
	\caption{\textbf{Comparison of the dHvA oscillation from magnetic torque and magnetization.}
	The oscillatory components of the magnetic torque (blue curve) measured using a piezo cantiliver and the magnetization (orange curve) measured using MPMS are plotted versus 1/$\mu_0$$H$. For both measurements, the magnetic field is along the $c$-axis of the sample. There is close agreement between the phases of the oscillations for both measurements.}
	\label{fig:Mcompare}
    \end{center}
\end{figure}

 \begin{figure}[!tht]
 \renewcommand{\thefigure}{S\arabic{figure}}
    \begin{center}
 \includegraphics[width=1\linewidth]{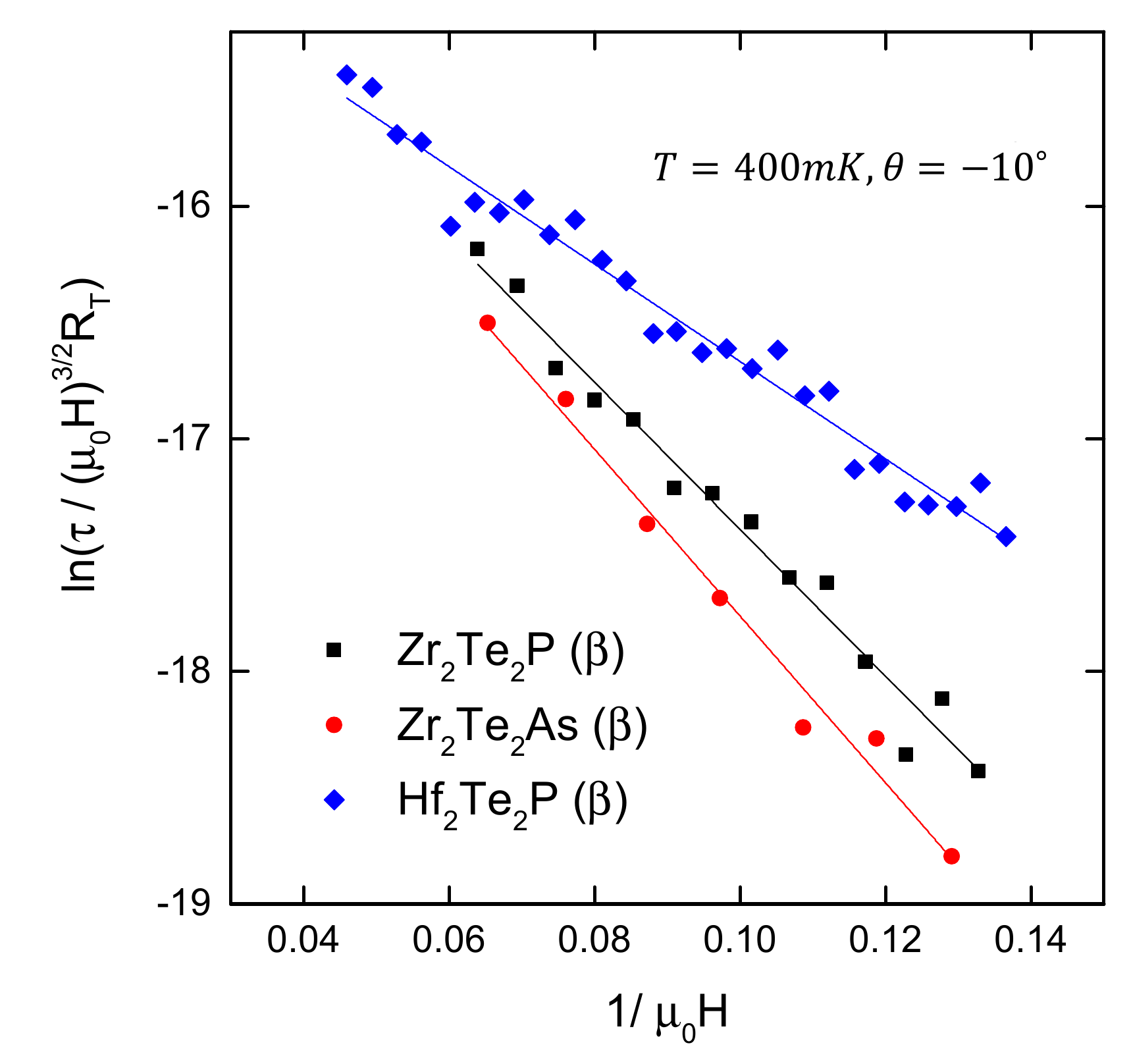}
	\caption{\textbf{``Dingle" plots of $\beta$ pockets of $T$$_2$$Ch$$_2$$Pn$ at $\theta$ = -10$^{\circ}$}. The Dingle temperatures $T_{\rm{D}}$ are determined using the straight line fits, as described in the text. $T_{\rm{D}}$ = 11 K for Zr$_2$Te$_2$P, $T_{\rm{D}}$ = 23 K for Zr$_2$Te$_2$As and $T_{\rm{D}}$ = 6.5 K for Hf$_2$Te$_2$P, indicating varying sample amounts of disorder scattering.}  
	\label{fig:dingle}
    \end{center}
\end{figure}

\begin{figure}[!tht]
\renewcommand{\thefigure}{S\arabic{figure}}
    \begin{center}
        \includegraphics[width= 0.8\linewidth]{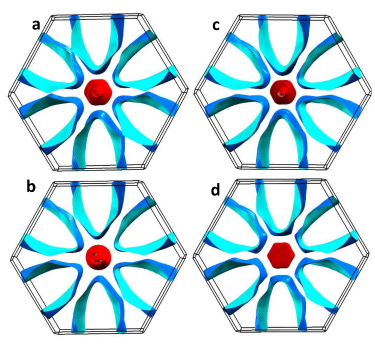}
       \caption{\textbf{Comparison between Fermi-surfaces calculated for all four tetradymite compounds}:
       Calculated Fermi surfaces for (a) Zr$_2$Te$_2$P (b) Zr$_2$Te$_2$As (c) Hf$_2$Te$_2$P (d) Ti$_2$Te$_2$P.
       The red (blue) sheets are hole-like (electron-like) pockets. While the Fermi surfaces in (a), (b), and (c) have the same number of hole pockets,
       in Ti$_2$Te$_2$P the smallest hole pocket around the $\Gamma-$point is missing.}
        \label{fig:DFTFS}
    \end{center}
\end{figure}

\begin{figure}[!tht]
\renewcommand{\thefigure}{S\arabic{figure}}
    \begin{center}
        \subfigure[]{
            \includegraphics[width=\figwidth]{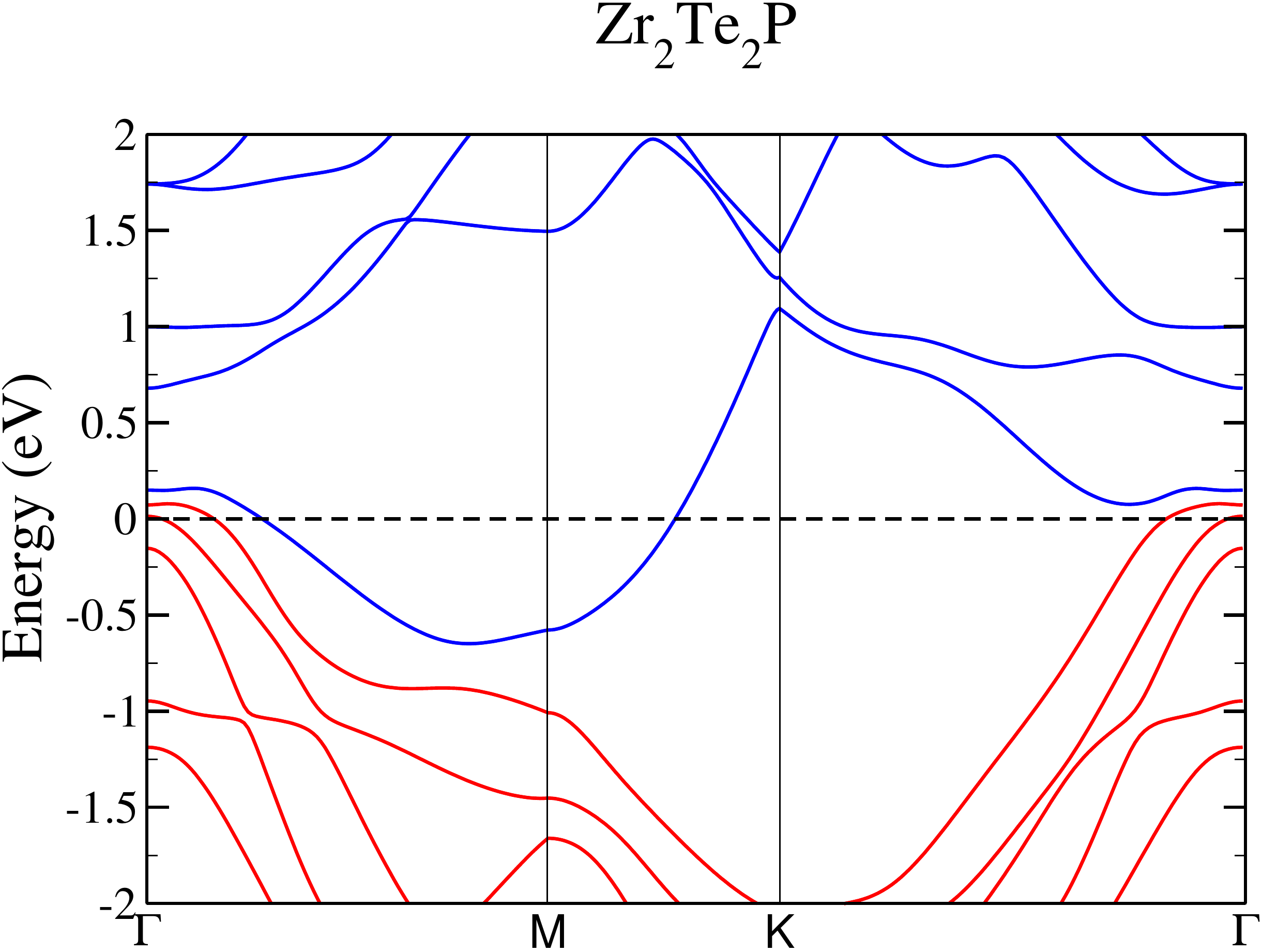}
            \label{fig:Zr2Te2P_rhombo}
        } \hskip 0.2 in
        \subfigure[]{
            \includegraphics[width=\figwidth]{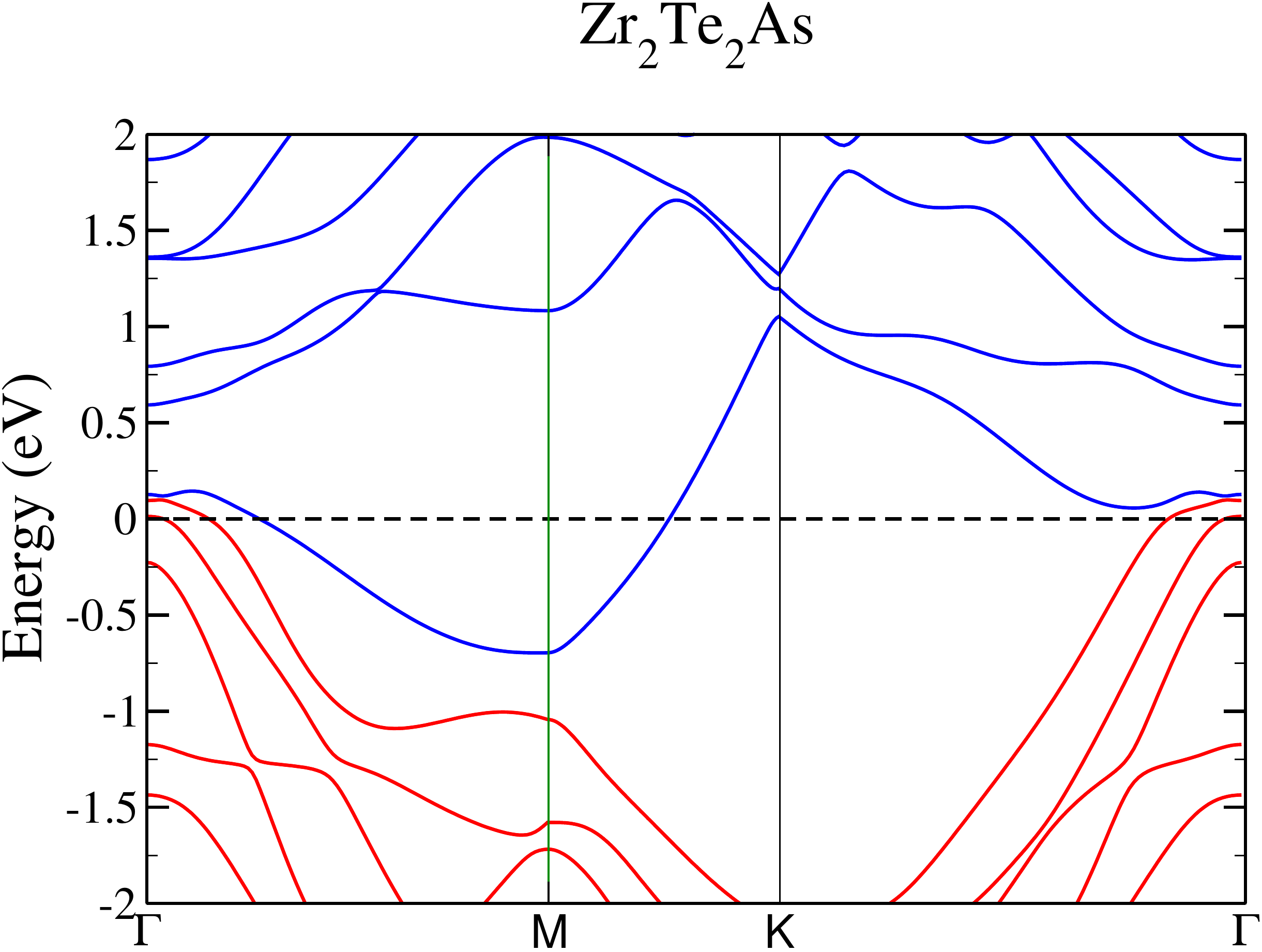}
            \label{fig:Zr2Te2As_rhombo}
        }
        \\
\vskip 0.1 in 
        \subfigure[]{
            \includegraphics[width=\figwidth]{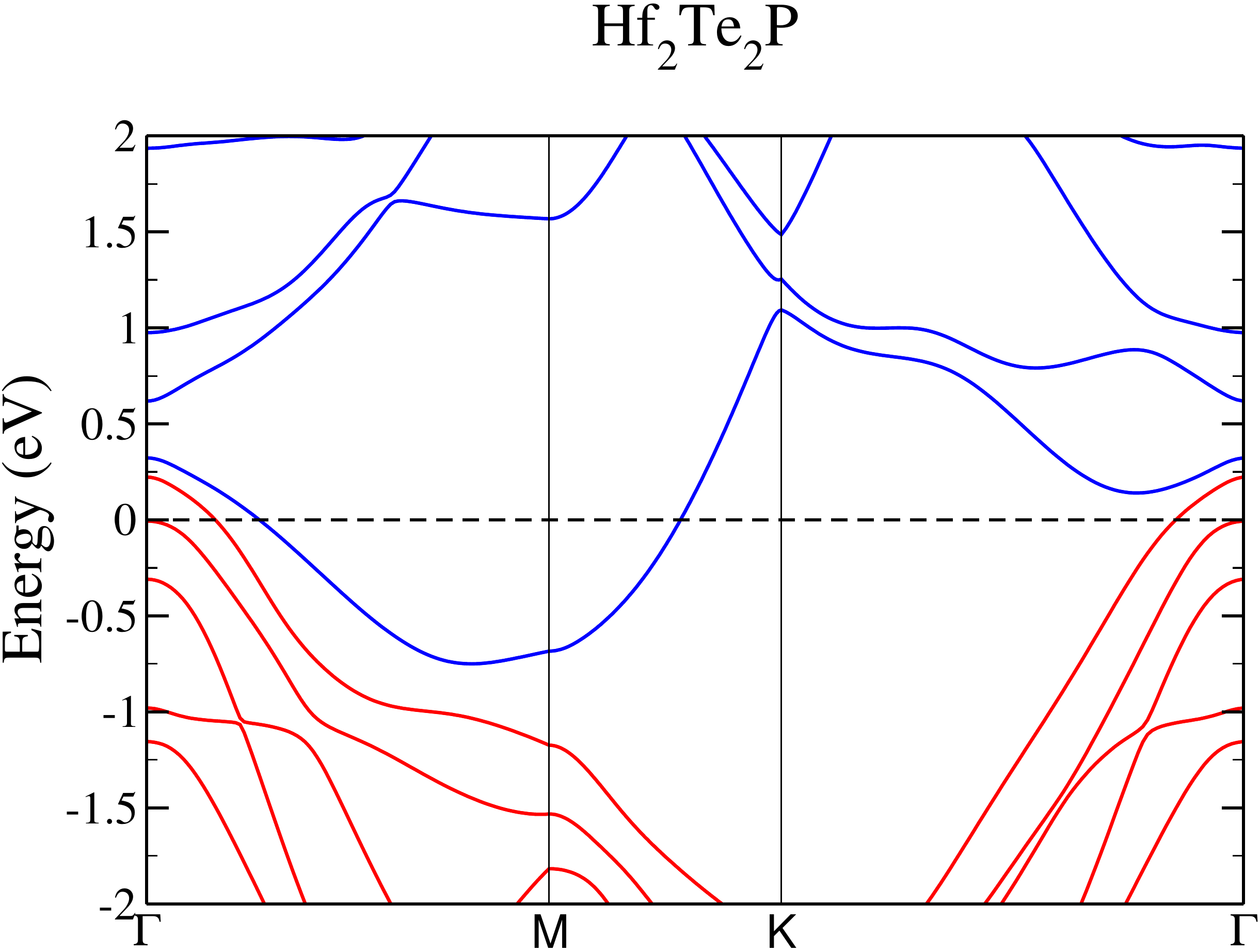}
            \label{fig:Hf2Te2As_rhombo}
        } \hskip 0.2 in
        \subfigure[]{
            \includegraphics[width=\figwidth]{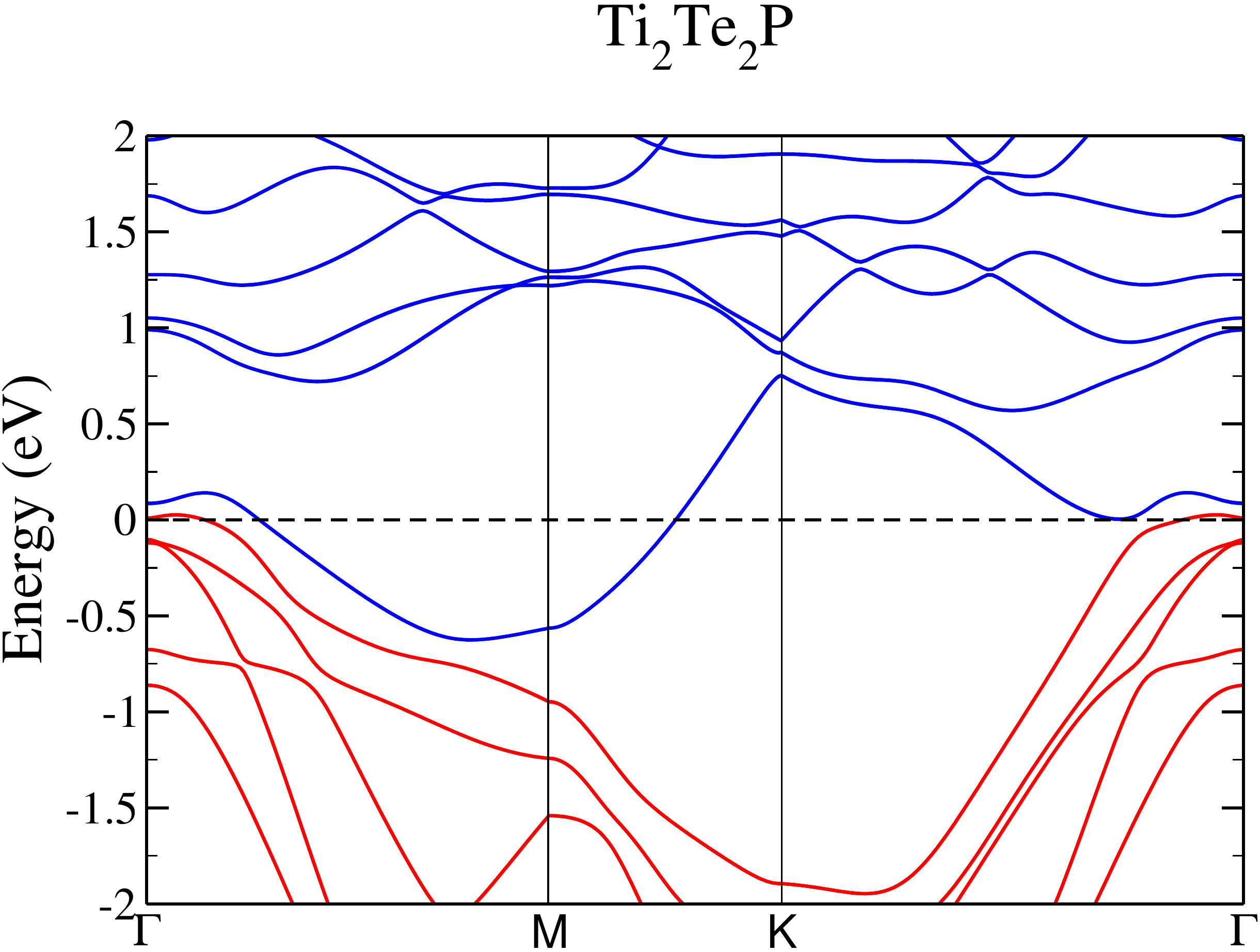}
            \label{fig:Ti2Te2As_rhombo}
        }
        \caption{\textbf{Comparison of the band-structure of the four crystals:} 
(a) Zr$_2$Te$_2$P (b) Zr$_2$Te$_2$As (c) Hf$_2$Te$_2$P (d) Ti$_2$Te$_2$P.
The bands corresponding to the electron (hole) pockets
are shown in blue (red).
}
    \label{fig:bands}
    \end{center}
\end{figure}

\begin{figure}[!tht]
\renewcommand{\thefigure}{S\arabic{figure}}
    \begin{center}
        \includegraphics[width= 0.6\linewidth]{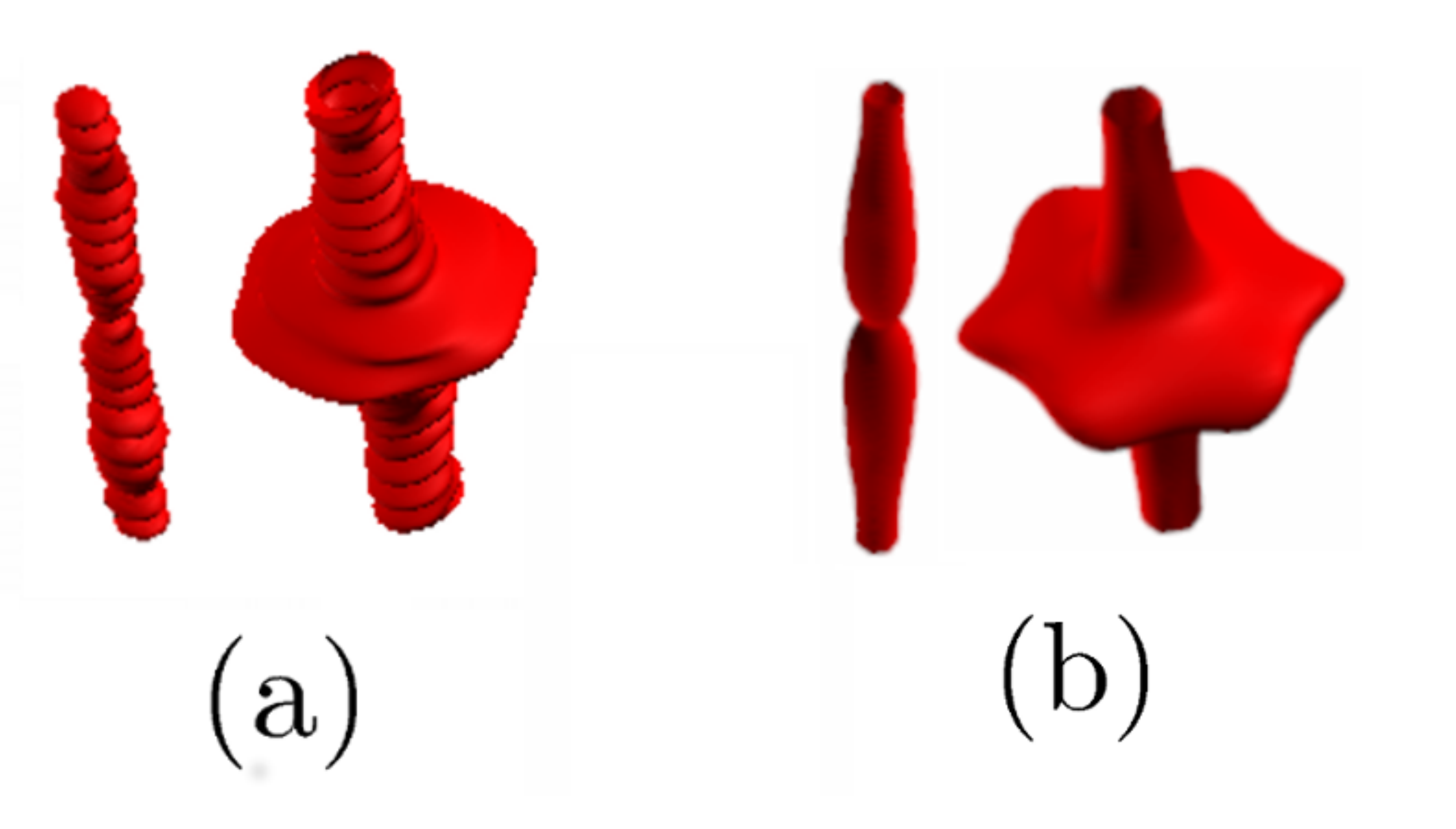}
 	\caption{\textbf{Comparison of the hole pocket in Zr$_2$Te$_2$P (a) before  and (b) after Wannierization}.
Notice that the wavy features of the hole pockets are no longer present 
after the Wannierization scheme where a dense k-mesh interpolation is used.
       }

        \label{fig:WannierHole}
    \end{center}
\end{figure}

\begin{figure}[!tht]
\renewcommand{\thefigure}{S\arabic{figure}}
    \begin{center}
        \includegraphics[width= 0.8\linewidth]{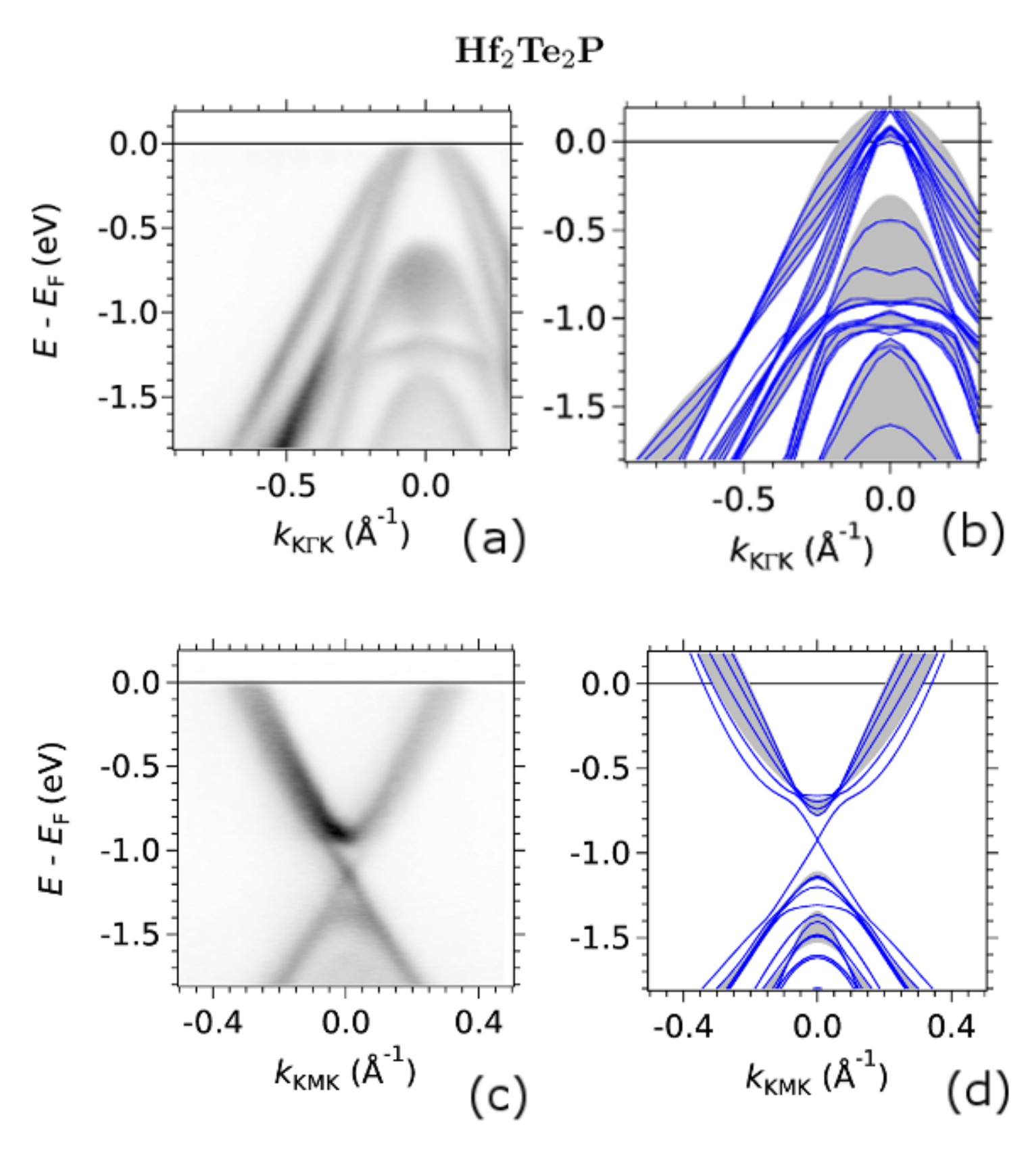}
 	\caption{\textbf{Comparison between ARPES measurements and band structure calculations for Hf$_2$Te$_2$P
for cuts along the $\bar{K}$ - $\bar{\Gamma}$ - $\bar{K}$ direction (upper panels) and along the $\bar{K}$ - $\bar{M}$ - $\bar{K}$ direction (lower panels).} 
(a) and (c) are ARPES determined band dispersion and (b) and (d) are DFT calculated bands.
Fermi energy $E_{\rm{F}}$ is indicated by the horizontal black line. Blue solid lines are bands obtained from a 5-layer slab calculation and therefore represent surface states. The grey ribbons represent the bulk bands.
       }
        \label{fig:HTPARPESDFT}
    \end{center}
\end{figure}

\begin{figure}[!tht]
\renewcommand{\thefigure}{S\arabic{figure}}
    \begin{center}
        \includegraphics[width= 0.4\linewidth]{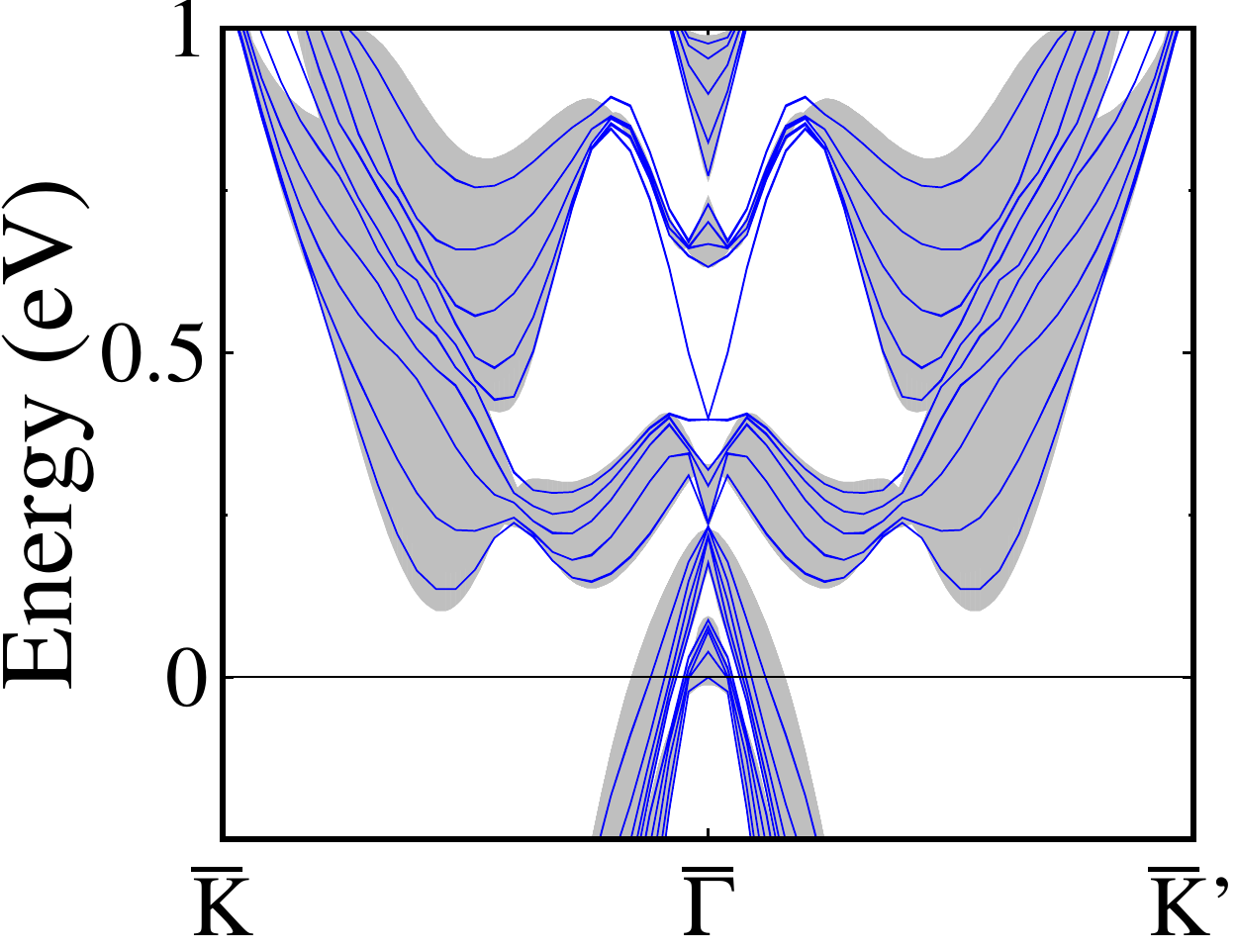}
       \caption{DFT calculated band dispersion for Hf$_2$Te$_2$P along $\bar{K}$ - $\bar{\Gamma}$ - $\bar{K}$ showing the presence of Dirac-like surface state at the $\bar{\Gamma}$ point.}
        \label{fig:HTPDFT}
    \end{center}
\end{figure}



\end{document}